\documentclass[useAMS, usenatbib, fleqn]{mn2e}
\pdfoutput=1
\usepackage{mathptmx}

\usepackage[T1]{fontenc}

\usepackage{aecompl}
\usepackage{amsmath}
\usepackage{amssymb}
\usepackage{gensymb}
\usepackage{wasysym}
\usepackage{graphicx}
\usepackage{subfig}
\usepackage{array}
\usepackage{url}
\usepackage{float}
\usepackage{xcolor}
\usepackage{hyperref}
\hypersetup{colorlinks=true,citecolor=blue,linkcolor=blue,filecolor=blue,urlcolor=blue}

\title[Galaxy selection and halo boundaries]
    {The impact of galaxy selection on the splashback boundaries of galaxy clusters}
\author[S.~O'Neil et al.]
    {{Stephanie O'Neil$^{1}$ \thanks{E-mail: sloneil@mit.edu},
    Josh Borrow$^{1}$,
    Mark Vogelsberger$^{1}$,
    and Benedikt Diemer$^{2}$}\\
    $^{1}$Department of Physics and Kavli Institute for Astrophysics and Space Research,
           Massachusetts Institute of Technology,
           Cambridge, MA 02139, USA \\
    $^{2}$Department of Astronomy, University of Maryland, College Park, MD 20742, USA\\
    }
\begin{document}

\date{Accepted 2022 March 21. Received 2022 March 18; in original form 2022 February 11}

\pagerange{\pageref{firstpage}--\pageref{lastpage}}
\pubyear{2022}

\maketitle
\label{firstpage}

\begin{abstract}
We explore how the splashback radius ($R_{\rm sp}$) of galaxy clusters, measured using the number density of the subhalo population, changes based on various selection criteria using the IllustrisTNG cosmological galaxy formation simulation.
We identify $R_{\rm sp}$ by extracting the steepest radial gradient in a stacked set of clusters in 0.5 dex wide mass bins, with our clusters having halo masses $10^{13} \leq M_{\rm 200, mean} / {\rm M}_\odot \leq 10^{15}$.
We apply cuts in subhalo mass, galaxy stellar mass, $i$-band absolute magnitude and specific star formation rate.
We find that, generally, galaxies of increasing mass and luminosity trace smaller measured splashback radii relative to the intrinsic dark matter radius.
We also show that quenched galaxies may be used to reliably reconstruct the dark matter splashback radius.
This trend is likely due to changes in the galaxy population.
Additionally, we are able to reconcile different observational predictions that $R_{\rm sp}$ based upon galaxy number counts and dark matter may either align or show significant offset (e.g. those using optically- or SZ-selected clusters) through the selection functions that these studies employ.
Finally, we demonstrate that changes in $R_{\rm sp}$ measured through number counts are not due to a simple change in galaxy abundance inside and outside of the cluster.
\end{abstract}


\section{Introduction}

Galaxy clusters are the largest objects in the Universe bound by their own self-gravity.
They are collections of hundreds to thousands of galaxies within a single dark matter halo along with stars and gas.
The environments within clusters differ significantly
from the field; they are sites of high density, are full of hot gas and host extreme physics like active galactic nuclei \citep[e.g.][and references therein]{Smith2005,Lim2021}.
This additionally impacts the galaxy population as they fall in to large haloes and evolve within these environments.
Galaxies in clusters are more likely to be stripped of gas, which quenches their star formation, and tend to be older and elliptical \citep[e.g.][and references therein]{Dressler1980,Cooper2006,Donnari2021}.
Understanding both the state of the cluster itself and the expected influence of its environment necessitates an understanding of the physical size of the cluster.
The physical extent of this environment informs our understanding of how the galaxy population evolves as they fall into clusters.

Common definitions of halo size are typically based on fixing the average density within the halo rather than basing them on physical properties of the halo.
$R_{\rm 200, mean}$ or $R_{\rm 500, crit}$, for example, set the average enclosed density to 200 times the mean density or 500 times the critical density of the Universe  respectively.
This definition for the cluster boundary, however, is often not large enough to fully reflect the dynamical nature of cluster halos.
Haloes often accrete matter beyond the virial radius, so using this to define halo size may not give an accurate picture of a halo's mass evolution \citep[e.g.][]{Cuesta2008}.
Additionally, infalling haloes begin to be stripped of their mass well outside of the virial radius \citep{Behroozi2014}, indicating that the cluster's influence expands beyond the virial radius and we should expect the galaxy population to begin to change beyond this radius as well.
Using a fixed overdensity mass and radius definition can also cause a halo to experience pseudoeveolution, or a change in the size of the halo due to the expansion of the Universe rather than a physical change of the halo itself \citep{Diemer2013}.

The so-called splashback boundary defines halo size relative to intrinsic halo properties by separating the infalling material from material already collapsed into the halo \citep{Diemer2021b}.
The region encompassed by orbiting material is where we expect the cluster environment to dominate and influence the properties of galaxies.
As material falls in to the halo, it passes through the halo and reaches its first apocenter where, in idealised spherical environments, it forms a caustic at the splashback radius, $R_{\rm sp}$ \citep[e.g.][]{Diemer2014,Adhikari2014,More2015,Shi2016a}.
In more realistic, non-spherical environments, this caustic is smoothed out but still manifests as a rapid change in the slope of the density profile.
This feature in the density profile can be used as an observational signature of the splashback radius.

$R_{\rm sp}$ has been studied extensively in numerical simulations. \citet{More2015} used the point of steepest slope in stacked halo density profiles as a proxy for $R_{\rm sp}$ and found that $R_{\rm sp}/R_{\rm200,mean}$ decreases with increasing accretion rate, a trend confirmed in hydrodynamic simulations by \citet{O'Neil2021}.
The reason for this is that the addition of new matter deepens the potential and thus shrinks the orbits of particles \citep{Adhikari2014}. \citet{Diemer2017b} traced the trajectories of particles as they fell into haloes in N-body simulations and found an additional, though weaker, dependence on halo mass and cosmology.
The radius of steepest slope does not exactly correspond to this dynamical definition \citep{Diemer2020a}, but it is less computationally intensive and more feasible in observations.

Although dark matter determines much of the halo dynamics, galaxy measurements present a more practical means of calculating the splashback radius.
However, it is not guaranteed that galaxies will exactly trace the dark matter distribution and provide the same splashback radius as dark matter measurements.
For example, \citet{Deason2020} studied Local Group simulations and found that $R_{\rm sp}$ measured from density profiles of satellite galaxies was significantly smaller than the dark matter $R_{\rm sp}$.
\citet{Xhakaj2020} also found a smaller $R_{\rm sp}$ measured for subhaloes in larger haloes of masses $M_{\rm 200, mean} \approx 10^{14}$ $\rm{M}_{\odot}$, while \citet{O'Neil2021} found that the subhalo results produced significantly smaller $R_{\rm sp}$ only for haloes less than $M_{\rm 200, mean} \approx 10^{13.5}$ $\rm{M}_{\odot}$ using a less massive population of subhaloes.
The difference between the subhalo and dark matter profiles is often attributed to dynamical friction \citep{Adhikari2016}, which would have a larger effect for less massive haloes and more massive subhaloes.

One aspect of the galaxy population that is naturally explained when considering the splashback radius are so-called backsplash galaxies.
These are galaxies that have passed through their host halo and are outside the virial radius, but they mostly remain within the splashback radius.
Understanding the details of this galaxy population matters because these galaxies are distinct from others at a similar location that have not yet entered the cluster, and their journey through the cluster typically strips them of a significant fraction of their mass \citep{Knebe2011}.
Although they may look similar to nearby galaxies, they have already experienced the dynamical effects of the cluster environment \citep{Gill2005, Pimbblet2011}, further complicating measurements made based on observational properties.

 Recently, there have been a number of detections of the splashback feature in observational surveys through the use of galaxy number counts. \citet{More2016} found a distinctive splashback feature in SDSS survey data, although the cluster selection used resulted in a much smaller splashback radius than expected \citep{Busch2017,Sunayama2019,Murata2020}.
 Using DES data, \citet{Baxter2017} found a distinct change in galaxy colour from blue to red at the splashback boundary, with \citet{Nishizawa2018} confirming this with CAMIRA clusters selected from HSC data.
\citet{Murata2020} used the same catalogue to show that red galaxies most reliably trace the splashback radius since they better represent the orbiting population than bluer galaxies. 

Other studies using SZ-selected clusters, namely \citet{Shin2019}, \citet{Zurcher2019}, \citet{Adhikari2021} and \citet{Shin2021}, found splashback radii closer to those predicted by numerical simulations.
Using weak lensing profiles, \citet{Contigiani2019} were able to constrain $R_{\rm sp}$ using a simple parametric model.
\citet{Chang2018} used galaxy number counts along with weak lensing to find that there was good agreement between $R_{\rm sp}$ measured from the underlying potential and number counts.
Additional studies by \citet{Bianconi2021} were able to constrain the splashback radius of clusters using the Local Cluster Substructure Survey.

Since these studies are only able to make measurements of observational components of haloes, e.g. galaxies, it is important to understand how these components relate to the theoretical boundary and any biases that may be present.
In addition, various surveys may be sensitive to different types of galaxies, so it is important to understand the impact this may have on the accuracy of the measurement of $R_{\rm sp}$.
Especially because the cluster galaxy population differs from the field galaxy population, we can expect that different types of galaxies may trace the splashback feature in different ways.
For instance, \citet{Adhikari2021} found that density profiles of red galaxies showed a distinct splashback feature while the density profiles of blue galaxies did not.
\citet{Dacunha2021} also showed different galaxy populations in simulations trace the splashback feature differently, with blue galaxies showing a shallower splashback feature and more massive red galaxies showing a splashback feature at a smaller radius.

With the progression of simulations \citep{Vogelsberger2020}, they have become a prime tool for understanding the relationship between observable properties of galaxies and the underlying properties of dark matter.
This work focuses on measuring the splashback feature of various populations of galaxies in large haloes in the IllustrisTNG simulations and comparing this to the splashback radius measured from the dark matter.
Using the methods developed in \citet{O'Neil2021}, we measure the point of steepest slope in the number density profile of galaxies of varying total mass, stellar mass and magnitude and compare this measurement to the point of steepest slope in the dark matter density profile.

The rest of the paper is structured as follows.
In Section \ref{sec:methods}, we describe the IllustrisTNG simulations, our halo and galaxy sample selection methods and our method for computing density profiles and identifying $R_{\rm sp}$.
We compare measurements of $R_{\rm sp}$ for different galaxy populations and dark matter in Section \ref{sec:results}.
In Section \ref{sec:discussion}, we discuss how the change in galaxy properties may impact the measurements of the splashback radius.
Finally, we summarise our conclusions in Section \ref{sec:conclusions}.

\section{Methods}
\label{sec:methods}

\begin{figure*}
    \centering
    \includegraphics{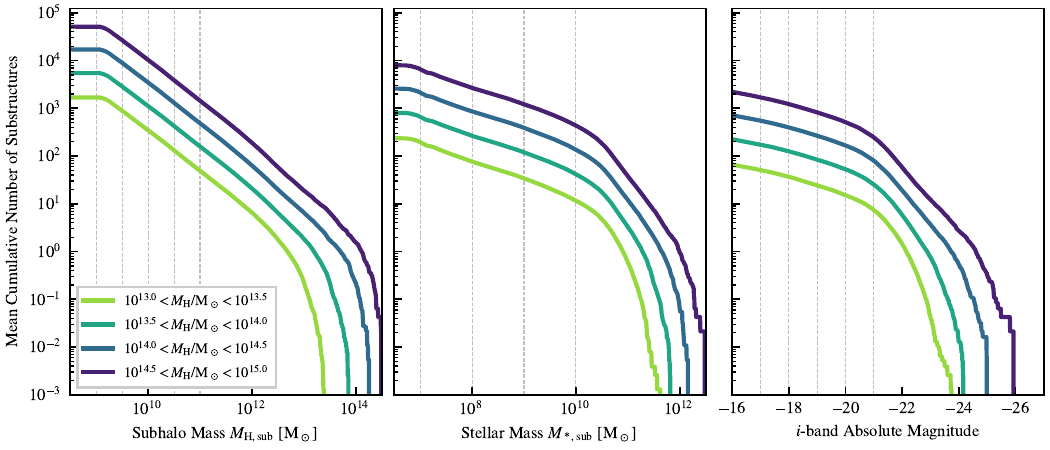}
    \caption{The distribution of subhaloes in our three initial levels of cuts: left panel shows the halo mass distribution of substructures, the central panel shows the stellar mass distribution, and the right panel shows the $i$-band magnitudes of all substructures, split by cluster halo mass. These are cumulative from the highest mass (or brightest) end, meaning that the lines show the average number of galaxies per cluster that would be included in a cut at that independent variable. The cuts used in the rest of the paper are denoted by dashed grey vertical lines.}
    \label{fig:halo_distributions}
\end{figure*}

\subsection{Simulations}
\label{sec:methods_simulations}

This work uses data from the IllustrisTNG simulation suite as described in \citet{Nelson2018, Marinacci2018, Springel2018,Naiman2018, Pillepich2018b}, which uses methods and physics updated from the original Illustris simulations \citep{Vogelsberger2014b}.
The simulations use a cosmology with $\Omega_{\rm m} = \Omega_{\rm{dm}}+\Omega_{\rm{b}} = 0.3089,\ \Omega_{\rm{b}} = 0.0486,\ \Omega_{\Lambda} = 0.6911,\ \sigma_8=0.8159,\ n_s=0.9667$, and Hubble constant $H_0=100h\,\rm{km} \,\rm{s}^{-1}\,\rm{Mpc}^{-1}$ where $h=0.6774$ as given in \citet{PlanckCollaborationXIII2016}.
The suite consists of dark matter only and full hydrodynamic runs at various resolutions in box sizes with side lengths of 50 Mpc, 100 Mpc and 300 Mpc.
For this work, we use the largest volume box, TNG300-1, with side length 300 Mpc, to give us an adequate sampling of large haloes.
The full physics variant with hydrodynamics is used so we can identify subhaloes with baryonic properties to define our galaxy samples.

The simulations employ a tree and particle mesh (tree-PM) gravity method, with a moving mesh (finite volume) technique for magnetohydrodynamics, using the \textsc{Arepo} code \citep{Springel2010,Weinberger2020}.
TNG300-1 has $2500^3$ gas cells and $2500^3$ dark matter particles with a target cell mass of $1.1\times 10^7$  $\rm{M_{\odot}}$ and a dark matter particle mass of $5.9\times10^7$ $\rm{M_{\odot}}$.
For dark matter particles, the gravitational softening length is $1.5$ kpc in physical units for $z \leq 1$ and comoving units for $z > 1$.
The gas cells have an adaptive comoving softening length with a minimum of $0.37$ kpc.

Gas cells cool radiatively and through metal line cooling and then form stars stochastically following a two-phase effective equation of state \citep{Springel2003}.
Mass, metals and energy are returned through asymptotic giant branch stellar winds and supernovae.
The galaxy formation model uses a supernova wind model \cite{Pillepich2018a} and a radio mode active galactic nucleus feedback scheme \citep{Weinberger2017} updated from the original Illustris project \citep{Vogelsberger2014b} along with further numerical refinements \citep{Pakmor2016}.
Black holes are seeded in haloes starting at $1.2\times10^6\rm$ M$_{\odot}$ and can reach a mass of $7.4\times10^{10}$ $\rm{M}_{\odot}$ following \citet{DiMatteo2005}.
They grow by accreting gas following an Eddington limited \citet{Bondi1952} prescription and through mergers with other black holes.
Feedback is injected into the environment in a quasar or kinetic mode depending on the black hole's accretion rate as described in \citet{Weinberger2017}.

Haloes are identified within the simulation using a Friends-of-Friends (FoF) algorithm \citep{Davis1985}.
Particles are linked to each other when they lie within a linking length of $b=0.2$.
The FoF algorithm links both particles and gravitationally bound structures identified using \textsc{SubFind} \citep{Springel2001,Dolag2009}.
The most massive gravitationally bound object in a FoF group is labelled as the main halo while the others within the group are identified as subhaloes.
The centre of the halo is defined as the position of the most bound particle of the main halo.

Synthetic magnitudes are generated in four bands, $g$, $r$, $i$ and $z$, corresponding to the SDSS Camera Response Function (with an airmass of 1.3).
Details on these specific filters can be found in \citet{Stoughton2002} section 3.2.1.
Here, we use the $i$-band magnitude as this corresponds to the peak emission of the selected galaxies at $z=0$.

\subsection{Halo selection}

We follow the same halo sample selection as in \citet{O'Neil2021}, which we summarise here.
Over a redshift range of $0\leq z\leq0.5$, we take haloes with $10^{13}
\leq M_{\rm200,mean}/\rm{M}_{\odot} \leq 10^{15}$.
We remove from this sample haloes that are within $10R_{\rm 200, mean}$ of a more massive halo to ensure that haloes in our sample are not being disrupted by a more massive object.
This gives us 1401 halos at $z=0$ in the TNG300-1 simulation. Here, we take $R_{\rm 200, mean}$ to be the radius that encloses matter a density equal to 200 times the density of the Universe.

\subsection{Galaxy definition}
\label{sec:methods_gal_def}

Our fiducial galaxy definition is modified from \citet{O'Neil2021}.
All substructure identified by the galaxy finder (32 particles or more) is included in our base sample, to which we apply different cuts. Additionally, we include galaxies in the radius range $0.1 R_{\rm 200, mean} < r < 10.0 R_{\rm 200, mean}$, modified from the maximum radius of $5.0 R_{\rm 200, mean}$ used in \citet{O'Neil2021}.
This is to aid with the fitting of the outer density profile (see the later discussion on identifying $R_{\rm sp}$, and Appendix \ref{app:fitting}).

In this paper, we explore the impact of varying the sampling of substructure on the measurement of the splashback feature.
We therefore make measurements of the splashback feature using subsets of this galaxy population by varying a cut in total mass, stellar mass and absolute $i$-band magnitude.
The average cumulative number of galaxies included in a cut for each halo are shown in Figure \ref{fig:halo_distributions} as a function of their halo mass, stellar mass and $i$-band absolute magnitude.






\subsection{Density profiles}
\label{sec:methods_density_profiles}

To construct a density profile, we stack profiles from main haloes with similar masses in four bins with edges $M_{\rm200, mean}/\rm{M}_{\odot} =$ $10^{13.0}$, $10^{13.5}$, $10^{14.0}$, $10^{14.5}$ and $10^{15.0}$.
The number of haloes in each mass ranged at select redshifts is summarised in Table \ref{table:sample_size}.

\begin{table}
\centering
\begin{tabular}{c|cccc}
     Redshift $z$ & \multicolumn{4}{c}{$\log_{10}\left(M_{\rm 200, mean}\,/\,{{\rm M}_{\odot}}\right)$} \\
     & \!{13.0-13.5}\! & \!{13.5-14.0}\! &  \!{14.0-14.5}\! & \!{14.5-15.0}\! \\
     \hline
     $0.0$  & 770 & 394 & 185 & 47 \\
     $0.5$  & 991 & 444 & 119 & 13
\end{tabular}
\caption{The number of cluster haloes identified for use in this work found in TNG300-1 for each mass range at select redshifts.}
\label{table:sample_size}
\end{table}

To get the dark matter profile, we spherically average the dark matter particles in 128 radial bins logarithmically spaced between $0.01R_{\rm200,mean}<r<5.0R_{\rm200,mean}$.
We calculate the density profile for each halo and then take the mean value for haloes within each mass range in each radial bin.
Errors on the dark matter profiles are given as the standard deviation of the density profiles constituting the stack. 

For the galaxy profiles, we use 32 logarithmically spaced radial bins between $0.01R_{\rm200,mean}<r<10.0R_{\rm200,mean}$.
We then sum the number of galaxies in each bin and divide by the total volume of the radial bins across all haloes in a given mass range.
We repeat this calculation for each of our galaxy definitions as described in Section \ref{sec:methods_gal_def}.
This differs from \citet{O'Neil2021}, who use 42 linear radial bins and the median profile, since the reduced sample size of the subhalo population due to the subsampling in this work increased the noise in the subhalo profiles (see Appendix \ref{app:binning}).

Error bars on the number density profiles and gradients are given through bootstraps (re-sampling with replacement) of the individual clusters in the stack.
For an individual profile, 32 bootstraps are used to calculate the 16-84 percentile range and median for use in the fitting process.
We use such a low number of bootstraps here as these errors are only used to inform the fitting procedure, and any measurements of the splashback radius use 2048 top-level bootstraps (so in total 65536 bootstraps for a given measurement).

\subsection{Identifying $R_{\rm sp}$}
\label{sec:methods_rsp}

\begin{figure*}
    \centering
    \includegraphics{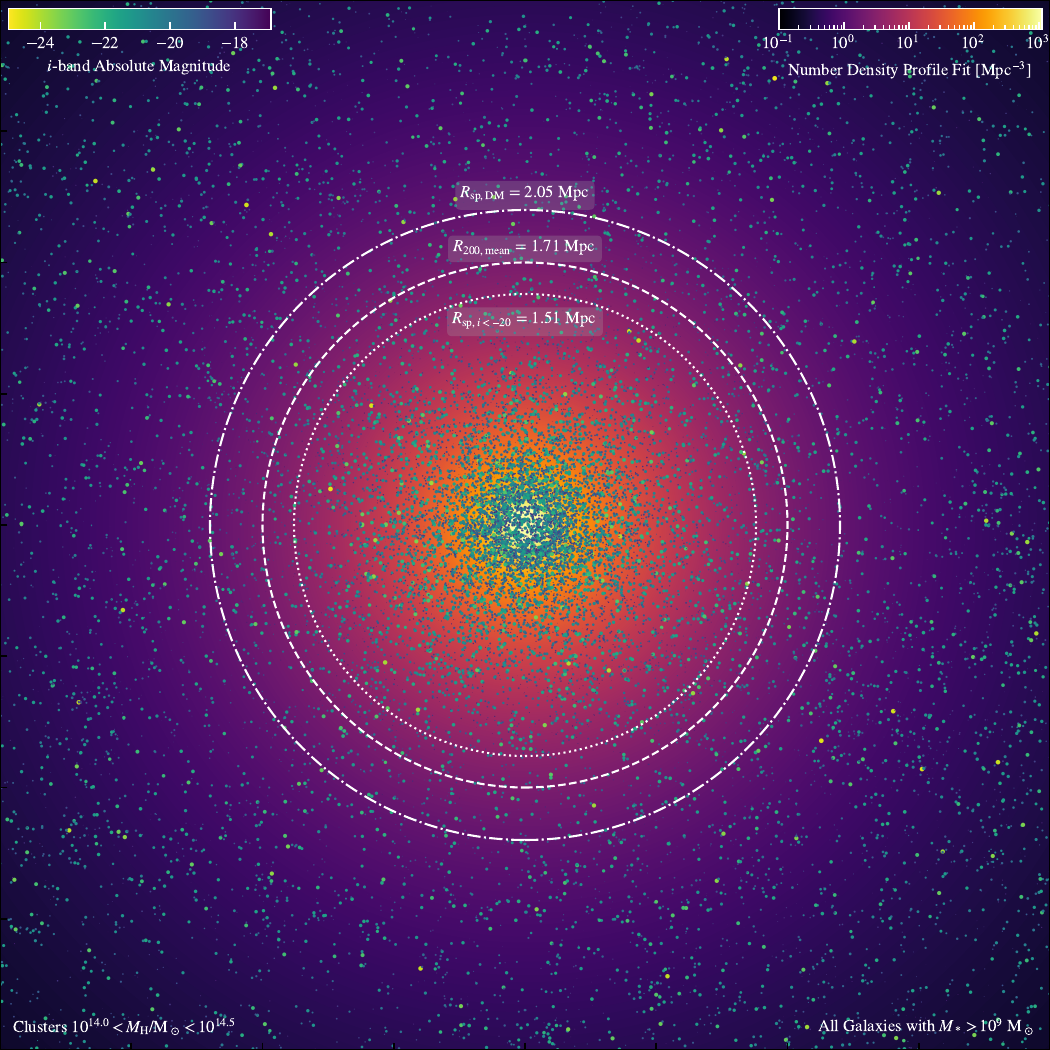}
    \caption{The combined cluster galaxy population (with $M_* > 10^9$, excluding the central 0.1 $R_{\rm 200, mean}$) for all stacked $10^{14.0} < M_{H} / {\rm M}_\odot < 10^{14.5}$ clusters. Each point shows an individual cluster galaxy sized by the logarithm of its stellar mass and coloured by its $i$-band magnitude. In the background, the density fit to this galaxy population is shown. Three circles show representative radii for this cluster: the dark matter splashback radius $R_{\rm sp, DM}$, $R_{\rm 200, mean}$ and the splashback radius measured from galaxies with $i$-band absolute magnitude below $-20$, $R_{{\rm sp}, i<-20}$. If using all galaxies in this figure, the splashback radius $R_{{\rm sp}, M_* > 10^9}$ is measured to be approximately $R_{\rm 200, mean}$.}
    \label{fig:gal_splashback_comparison}
\end{figure*}

\begin{figure}
    \centering
    \includegraphics{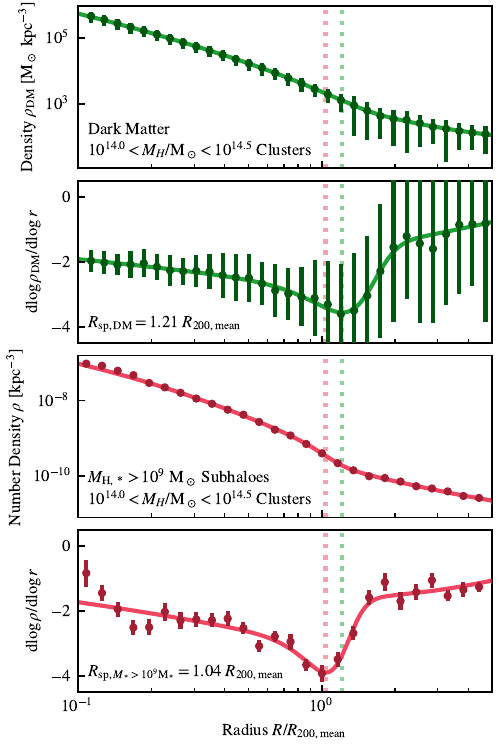}
    \caption{The dark matter (top two panels) and galaxy (bottom two panels) density and density gradient (top and bottom respectively) profiles for the cluster mass range $10^{14.0} < M_H / {\rm M}_\odot < 10^{14.5}$. Error-bars for the galaxy profile are generated through bootstraps of the profile generation; clusters are re-selected with replacement and for the galaxies we show a single `selection A' bootstrap (see Appendix \ref{app:fitting}). For the dark matter, the error bars represent the variance amongst individual profiles in this sample. Fits to Eqn. \ref{eq:density_derivative} are shown as solid lines in the gradient, with the same parameters used to demonstrate the associated profile in the density profile panels. The splashback radii yielded from the fitting process is shown with appropriately coloured dashed lines throughout all panels.}
    \label{fig:dm_gal_profiles}
\end{figure}

We identify $R_{\rm sp}$ as the minimum of the logarithmic slope of the density profile.
We calculate this point by fitting our density profiles to the function proposed in \citet{Diemer2014}:
\begin{align}
\label{eq:density}
    \indent\rho(r) &= \rho_{\rm{inner}} \times f_{\rm{trans}} + \rho_{\rm{outer}} \\
    \indent\rho_{\rm{inner}} &= \rho_{\rm{Einasto}} = \rho_{\rm{s}} \exp{\left( -\frac{2}{\alpha} \left[\left(\frac{r}{r_{\rm{s}}}\right)^{\alpha} - 1 \right] \right)}  \nonumber \\
    \indent f_{\rm{trans}} &=  \left[ 1 + \left(\frac{r}{r_{\rm{t}}}\right)^\beta \right]^{-\frac{\gamma}{\beta}} \nonumber \\
    \indent\rho_{\rm{outer}} &= \rho_{\rm{m}} \left[b_e \left( \frac{r}{5R_{200\rm{m}}} \right)^{-S_e} +1 \right]\:. \nonumber
\end{align}
Rather than fit to the density profile then take the derivative of the fit, we fit the derivative of Equation \ref{eq:density}, given in Equation \ref{eq:density_derivative}, to the numerical derivative of our density profiles, which can identify the splashback feature more robustly \citet{O'Neil2021}.

\begin{align}
    \label{eq:density_derivative}
    \indent & \frac{d\log{\rho}}{d\log{r}} = \frac{r}{\rho} \frac{d\rho}{d r} \\
    {\rm with}\nonumber\\
    &\frac{d\rho}{d r} = \frac{d\rho_{\rm inner}}{dr} \times f_{\rm trans} + \rho_{\rm inner} \times \frac{df_{\rm trans}}{dr} + \frac{d\rho_{\rm outer}}{dr}\:, \nonumber
\end{align}
where $\rho_{\rm inner}$, $f_{\rm trans}$, and $\rho_{\rm outer}$ are given in Equation \ref{eq:density} with derivatives
\begin{align}
    \label{eq:function_derivatives}
    \indent\frac{d\rho_{\rm inner}}{dr} &=  -\frac{2}{r_s} \left(\frac{r}{r_s}\right)^{\alpha-1} \times \rho_{\rm inner}  \nonumber \\
    \frac{df_{\rm{trans}}}{dr} &=  \left(1 + \left(\frac{r}{r_t}\right)^{\beta}\right)^{-\frac{\gamma}{\beta} - 1} \left(-\frac{\gamma}{r_t}\right) \left(\frac{r}{r_t}\right)^{\beta - 1} \\
    \frac{d\rho_{\rm{outer}}}{dr} &= -\frac{\rho_{\rm m} b_e s_e}{5R_{\rm 200m}} \left(\frac{r}{5R_{\rm 200m}}\right)^{-s_e - 1}\:. \nonumber
\end{align}

This function combines descriptions for the inner region of a halo using the Einasto profile for $\rho_{\rm inner}$, a transitional region $\rho_{\rm trans}$, and an outer region $\rho_{\rm outer}$ where the profile flattens to the mean density of the Universe, $\rho_{\rm m}$.
The parameters $\rho_s, r_s, r_t, \alpha, \beta, \gamma, b_e,$, $S_e$, and $\rho_{\rm m}$ are left free to vary in our fits.
Errors on splashback radii measurements are calculated by re-sampling the cluster population with replacement 2048 times and calculating the median and 16-84 percentile range.
More details on the specifics of this fitting procedure are available in Appendix \ref{app:fitting}.

Figure \ref{fig:gal_splashback_comparison} shows an example fit, visually demonstrating the substructures that are used as part of the fit in projection and a number of relevant radii for this stack.
The galaxy population using subhaloes with a stellar mass greater than $10^9\rm{M}_{\odot}$ for all clusters in the mass range $10^{14.0-14.5}\rm{M}_{\odot}$ is shown with dots coloured by $i$-band magnitude and sized by subhalo radius.
The background is coloured by the mean subhalo number at that radius for the set of haloes.
The number density profile of galaxies is fit for subhaloes with $i<-20$, and $R_{\rm sp, i}$ is identified as described in Section \ref{sec:methods_rsp} and shown with the dotted line.
Using the mean dark matter density for each halo in the same mass range, we fit the dark matter profile with Equation \ref{eq:density_derivative} and identify $R_{\rm sp,DM}$, shown with the dotted-dashed line.
$R_{\rm200 mean}$ is shown with the dashed line.

The density profiles and gradients along with the fit using Equations \ref{eq:density} and \ref{eq:density_derivative} are shown in Figure \ref{fig:dm_gal_profiles}.
The dark matter profile and gradient are shown in green in the top two panels, and the galaxy profile and gradient are shown in red in the bottom two panels.
The points show the mean value for the dark matter and galaxy density and gradient in each radial bin.
Error bars on the points for the galaxy profile are obtained through bootstrapping as described in Section \ref{sec:methods_density_profiles}.
For the dark matter, the errorbars represent the variance amongst the sample of haloes.
The analytic fits (Equations \ref{eq:density} and \ref{eq:density_derivative}) are shown with the solid lines.
The dotted lines through all for panels show the splashback radius for the dark matter (green) and galaxies (red), identified at the minimum of the gradient for each fit.
The splashback radius for the galaxies occurs at a slightly smaller radius than for the dark matter, which is typical of our galaxy populations.

\section{Results}
\label{sec:results}

In this section, we show the impact of various subhalo total mass, stellar mass and $i$-band absolute magnitude cuts on the recovered splashback radius.
We compare $R_{\rm sp}$ calculated for each of these cuts using the point of steepest slope for that galaxy population and compare to the point of steepest slope of the dark matter profile for the same set of haloes, which are grouped by mass as described in Section \ref{sec:methods_density_profiles}.
We also explore the redshift evolution of the $R_{\rm sp}$ measurements and the impact of using quenched or active galaxies.

\subsection{Subhalo mass cuts}
\label{sec:results_subhaloes}

\begin{figure}
    \centering
    \includegraphics{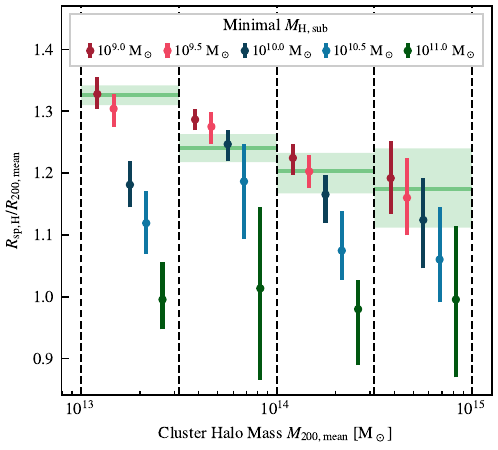}
    \caption{Shows how the splashback radius (relative to $R_{\rm 200, mean}$) varies as a function of cluster mass (bins are denoted with the vertical dashed lines, with the individual points displaced within the bin for clarity) and subhalo mass cut (different coloured points). Error bars show 16-84 percentile ranges across 2048 bootstraps, with the point showing the median value. The shaded regions and green bands show, for each cluster mass, the mean stacked dark matter splashback radius and its 16-84 percentile range. It is crucial to note that when using all intrinsic substructure ($M_{\rm H, sub} > 10^9$ M$_\odot$) we reproduce the dark matter splashback radii to within less than 1 sigma in the expected number of cases ($\approx$75\%), consistent with \citet{O'Neil2021}.}
    \label{fig:comparemcluster}
\end{figure}

\citet{O'Neil2021} showed that subhaloes in TNG-300 were able to capture the splashback feature.
To clarify the extent to which galaxies can reconstruct a splashback feature across a wide range of cluster masses, we show in Figure \ref{fig:comparemcluster} how different substructure mass cuts can lead to changes in the reconstructed splashback radius within each cluster mass bin.
We denote the splashback radius measured with subhaloes as $R_{\rm sp,H}$ and the splashback radius measured from the dark matter profiles as $R_{\rm sp,DM}$.
We sample the total mass of the subhaloes by using lower halo total mass (which includes particles bound to the strucutre and in the FoF group) limits of $10^{9}$ $\rm{M}_{\odot}$, $10^{9.5}$ $\rm{M}_{\odot}$, $10^{10}$ $\rm{M}_{\odot}$, $10^{10.5}$ $\rm{M}_{\odot}$, and $10^{11}$ $\rm{M}_{\odot}$.

Figure \ref{fig:comparemcluster} shows, for each cluster and subhalo mass range, the splashback radius relative to $R_{\rm200, mean}$.
The dark matter splashback radius, for the same set of haloes, is shown in green.
The solid line shows the measured value for our sample with the shaded region showing the estimated error on our calculation.
The error bars on $R_{\rm sp}$ show the $16^{\textrm{th}}$ and $84^{\textrm{th}}$ percentiles as the lower and upper bounds.
See Appendix \ref{app:fitting} for significantly more detail on this process.
While fits that include low-mass subhaloes tend to have $R_{\rm sp}$ in agreement with the dark matter calculations, only using higher mass subhaloes results in a measured $R_{\rm sp,H}$ significantly lower than $R_{\rm sp, DM}$.
For the cluster mass range with the least amount of evolution and well constrained error bars, $10^{14.0} < M_{\rm H} / {\rm M}_\odot < 10^{14.5}$, moving from structures with $M_{\rm H, sub} > 10^9$ M$_\odot$ (with $R_{\rm sp, H} \approx R_{\rm sp, DM}$) to $M_{\rm H, sub} > 10^{11}$ M$_\odot$ leads to a 20\% reduction in the measured splashback radius.

This trend is expected if dynamical friction is a primary source of the difference between $R_{\rm sp,H}$ and $R_{\rm sp,DM}$, \citep[e.g.][]{Adhikari2016,O'Neil2021}, although it has been claimed in the literature that tidal disruption and dynamical friction are not expected to play a role at splashback scales especially in the most massive clusters \citep[e.g.][]{Contigiani2019}.
Dynamical friction allows for the transfer of orbital energy and angular momentum to the host halo, reducing the orbital radius of the substructure, with the strength of this friction depending on the mass ratio of the substructure to the host, $M_{\rm H, sub} / M_{\rm 200, mean}$.
Dynamical friction is expected to be strong when the mass ratio exceeds a few percent \citep{Binney2008, Mo2010}.
More massive subhaloes experience more friction and lose more energy, and they therefore fall to a smaller radius relative to smaller subhaloes in the same host halo, which is consistent with our Figure \ref{fig:comparemcluster}.

While \citet{O'Neil2021} found a difference between the dark matter and galaxy splashback radius only for low-mass host haloes, however, we find that this difference persists across halo mass when subhaloes are also separated by mass.
It therefore seems that the difference between $R_{\rm sp,H}$ and $R_{\rm sp, DM}$ depends primarily on subhalo mass rather than the mass ratio between main halo and subhalo mass.
We also note that our study focuses on the current mass of the subhaloes while effects from dynamical friction may depend more on the mass at the time of accretion.
However, since more massive present-day subhaloes tend to come from more massive accreting subhaloes, we still expect any impact that dynamical friction has to increase with subhalo mass.
An additional consideration is that dynamical friction is believed to be significantly underestimated in relatively low-resolution cosmological simulations like TNG-300 \citep{vandenBosch2018, Morton2021}.

The splashback radius of subhaloes in simulations has also been explored in  \citet{Xhakaj2020} and \citet{Contigiani2021}.
\citet{Xhakaj2020} found a significantly smaller $R_{\rm sp}$ for cluster  haloes of $M_{\rm 200, mean} \approx 10^{14}$ $\rm{M}_{\odot}$ and subhaloes with mass $M_{\rm peak}\approx10^{12}$ $\rm{M}_{\odot}$, which exceeds the largest limit of our subhalo samples.
They found that the splashback radius measured from the subhaloes was roughly $10-15\%$ smaller than the dark matter splashback radius, although the exact difference depended on the accretion rate of the main halo.
This is similar to the difference we find for similar mass haloes and our largest subhalo mass cut.

Using zoom simulations of 24 massive clusters ($14.0 \leq M_{\rm 200, mean} / {\rm M}_\odot \leq 15.5$) \citet{Contigiani2021} found that splashback radii measured from the density of galaxies, substructure and dark matter are all consistent, independent of any applied mass cut for the substructure, diverging from our results.
Despite this in our most massive cluster bin, where we have 47 clusters at redshift $z=0$, we find that all results are consistent to within 1 $\sigma$ when applying various cuts in substructure mass.
Therefore we posit that it is likely that with an improvement in sample size a similar trend would be observed in C-EAGLE.

When using all substructure (with a substructure mass resolution of $M_{\rm H, sub} \approx 3 \times 10^8$ M$_\odot$), the dark matter only constrained realisation simulation SIBELIUS-DARK found that substructures trace an approximately 10\% smaller radius than the underlying dark matter \citep{McAlpine2022}.
The two measurements are, however, within 1$\sigma$ of each other and are hence still closely consistent with our predictions that low-mass haloes trace the dark matter.
This result was measured on single clusters (The Coma cluster with $M_{\rm 200, crit} = 1.3\times10^{15}$ M$_\odot$, and the Virgo cluster $M_{\rm 200, crit} = 3.5 \times 10^{14}$ M$_\odot$), a noticeable departure from our stacked measurements that may wash out small details.

\subsection{Stellar mass cuts}
\label{sec:results_galaxies}

\begin{figure}
    \centering
    \includegraphics{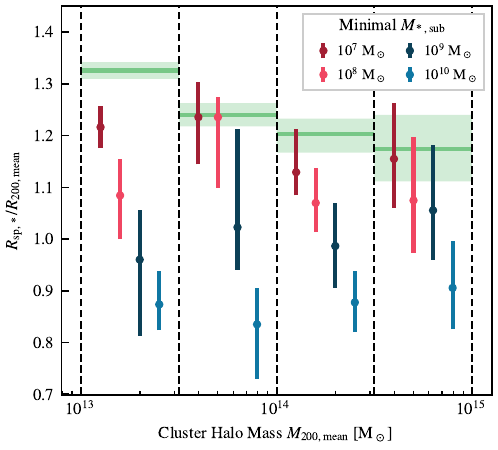}
    \caption{The analogue of Figure \ref{fig:comparemcluster}, but now using a cut in galaxy stellar mass instead of halo mass for each different colour. The green line again shows $R_{\rm sp, DM}$ measured from the underlying dark matter halo. Here we again see a significant trend where galaxies of higher stellar masses trace lower $R_{\rm sp}$, which is reproduced across the cluster mass range.}
    \label{fig:comparemclusterstellar}
\end{figure}

The properties of galaxies in clusters can vary significantly from galaxies in the field.
We now investigate how cuts in one of the most fundamental properties, stellar mass, affects the splashback feature.
The stellar mass of galaxies may be affected by various processes as they fall in to clusters, e.g. a burst of star formation followed by quenching \citep{Armitage2018}.
It is not clear, however, that this process occurs uniformly across the galaxy mass range.
As such, the stellar properties of cluster galaxies may trace different splashback radii than their host halo masses imply.
Additionally, this allows us to see the impact of only including visible substructure.

We vary the stellar mass cut with lower limits $M_* = 10^{7}$ M$_\odot$, $10^{8}$  M$_\odot$, $10^{9}$  M$_\odot$, and $10^{10}$  M$_\odot$ without regard to total subhalo mass or magnitude.
Here, we use the total stellar mass bound to the FoF subgroup to represent the stellar mass.
Figure \ref{fig:comparemclusterstellar} shows the splashback radius for each of our halo mass ranges.
The green lines give the median measurement for the dark matter profile of that halo mass range while the coloured points show the median splashback radius for each of the galaxy stellar mass cuts as described above, offset within each cluster mass range for clarity.
The error bars again denote the $16^{\rm th}$ and $84^{\rm th}$ percentiles of 2048 bootstraps.
Similarly to the total mass of subhaloes, we see that higher stellar mass galaxies result in a smaller splashback radius for all main halo mass ranges.
Making a cut of galaxies with $M_* > 10^7$ M$_\odot$ allows for the reconstruction of splashback radii consistent with the measured $R_{\rm sp, DM}$ in all cases except our smallest cluster mass bin.

The cut in stellar mass of $M_* > 10^{10}$ M$_\odot$ is the first to reliably produce measured $R_{\rm sp, *} < R_{\rm 200, mean}$ across the mass range, a trend not seen in Figure \ref{fig:comparemcluster} even at high subhalo masses.
At the highest stellar masses, cluster galaxies typically have a stellar mass ratio $M_* / M_{\rm H, sub} \approx 0.02$  \citep{Armitage2018}, implying that a cut at $M_* = 10^{10}$ M$_\odot$ corresponds to a total subhalo mass of $M_{\rm H, sub} \approx 5 \times 10^{11}$ M$_\odot$.
This is just outside the range we studied using the total bound masses in Section \ref{sec:results_subhaloes}.
Calculating the splashback radius using a total mass cut of $M_{\rm H, sub} > 5 \times 10^{11}$ M$_\odot$, we recover $R_{\rm sp, H} = 0.92^{+0.21}_{-0.12} R_{\rm 200, mean}$ for the clusters in the mass range $10^{14.0} \leq M_{\rm 200, mean} / {\rm M}_\odot \leq 10^{14.5}$, consistent with the prediction of $0.88$ from the corresponding stellar mass cut.

\citet{Deason2020} calculated $R_{\rm sp}$ in Local Group simulations using both dark and luminous subhaloes (subhaloes with at least one stellar particle), where subhaloes had bound dark matter masses greater than $10^{7.3}$ M$_{\odot}$.
When calculating $R_{\rm sp}$ with dark subhaloes, they found that their measurements aligned with their dark matter $R_{\rm sp}$ at $R_{\rm sp}\sim1.4R_{\rm200,mean}$.
However, their luminous subhalo population, which was slightly more massive than their dark subhalo population, produced $R_{\rm sp}\sim0.6R_{\rm200,mean}$.
This smaller value for $R_{\rm sp}$ also corresponded to a second caustic in the dark matter profile, where particles reached their second apocenter.
This is consistent with our results that stellar cuts results in somewhat smaller $R_{\rm sp}$ especially in our low-mass halo sample.

Compared to the dark matter, stellar mass is affected significantly less by stripping as subhaloes fall in to the main halo; galaxies can lose roughly eight times more of their dark matter than their stellar matter \citep{Smith2016}.
The stability of the stellar mass of cluster galaxies can hence be used as a more accurate tracer of dynamical friction. Despite this, we do not see significant trends with the ratio $M_{\rm *, sub} / M_{\rm 200, mean}$, instead finding that the measured splashback radius is mainly dependent on absolute galaxy stellar mass. For instance, the measured splashback radius in the cluster mass range $10^{13.0} \leq M_{\rm 200, mean} / {\rm M}_\odot \leq 10^{13.5}$ is measured to the same as the $10^{14.0} \leq M_{\rm 200, mean} / {\rm M}_\odot \leq 10^{14.5}$ bin when using galaxies with $M_{\rm *, sub} > 10^{10}$ M$_\odot$.

\subsection{Luminosity Cuts}
\label{sec:results_luminosity}

\begin{figure}
    \centering
    \includegraphics{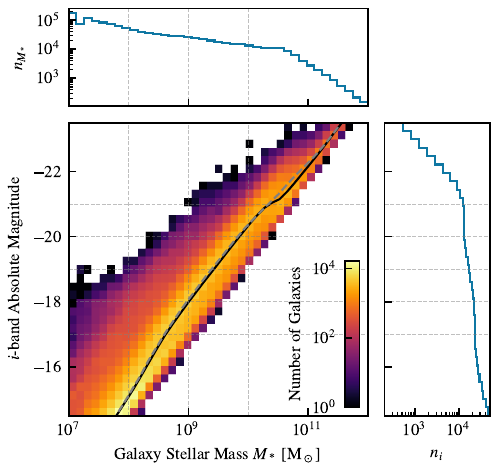}
    \caption{Stellar mass-absolute magnitude ($i$-band) relation in TNG-300 for the cluster galaxies in this sample. The horizontal and vertical dashed lines show the cuts made in Figures \ref{fig:comparemclusterstellar} and \ref{fig:iband_mcluster} for stellar mass and magnitude respectively. The solid black line shows the median relation using the stellar mass bins, and the grey dashed line shows the median relation using the $i$-band bins from the background histogram. The side panel histograms show the number of galaxies in each bin (note that these are not cumulatively stacked like Figure \ref{fig:halo_distributions}).}
    \label{fig:imagvsmstar}
\end{figure}

\begin{figure}
    \centering
    \includegraphics{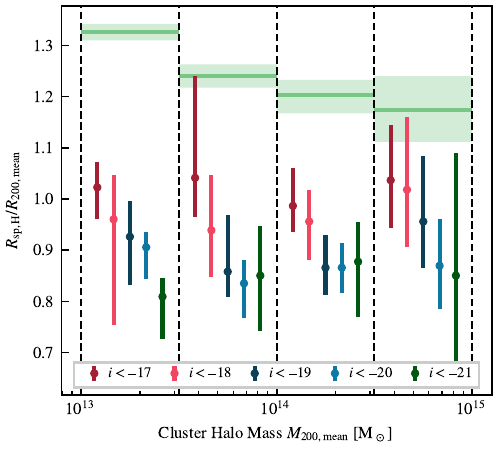}
    \caption{Another analogue of Figure \ref{fig:comparemcluster}, but now with each point representing the splashback radius with a cut in $i$-band magnitude. Note how the splashback radius decreases with increasing $i$-band maximum magnitude, meaning brighter cluster galaxies lead to smaller measured splashback radii.}
    \label{fig:iband_mcluster}
\end{figure}

Galaxy magnitude is a primary constraint in what many galaxy surveys can observe.
Although closely related to stellar mass, the relationship between magnitude and stellar mass is not exact.
Figure \ref{fig:imagvsmstar} shows this relationship for galaxies in our sample.
Notably, at a given stellar mass, there are a number of galaxies with magnitudes even 2 dex lower (brighter) than the median relation.
This implies that any cut in stellar mass will include a sub-population of significantly brighter galaxies than the the median relation suggests.
It also suggests that there will be a significant population of order-of-magnitude lower mass galaxies included in given $i$-band cut.
We therefore examine the impact of various $i$-band absolute magnitude limits on the measurement of $R_{\rm sp}$ similarly to the studies for total and stellar mass.

We divide our subhalo sample by upper $i$-band cuts of $-17$, $-18$, $-19$, $-20$ and $-21$.
This range is chosen such that many observational studies have magnitude cutoffs near the middle of this range.
For example, \citet{Baxter2017} uses $M_i-5\log(h)<-19.43$ and \citet{Shin2019}  uses $i<-19.87$.
Additionally, \citet{Zurcher2019} found that magnitude dependence of the splashback radius drops off for galaxies with $M_i-5\log h = -19.44$.
For IllustrisTNG, a value of $h=0.6774$ gives $-20.29$ for this limit, which is near the edge of our range of study.

Splashback radii extracted from stacked profiles using these cuts are shown in Figure \ref{fig:iband_mcluster}.
While we see a similar trend  as in Figures \ref{fig:comparemcluster} and \ref{fig:comparemclusterstellar} as lower magnitude (and hence higher mass) galaxies produce a smaller splashback radius, the magnitude cuts chosen all result in a significantly lower splashback radius than the dark matter splashback radius.

Our results are consistent with observational studies using optically selected clusters  that measure the splashback radius to be significantly smaller than what would be expected based on the underlying gravitational potential \citep[e.g.][]{More2016, Baxter2017,Nishizawa2018, Murata2020}.
In addition, \citet{Zurcher2019} found that $R_{\rm sp} \approx R_{\rm 200, mean}$ for minimal $i$-band absolute magnitudes $-19.4 < i_{\rm min} < -18.4$ using SZ-selected clusters.
This is broadly consistent with our findings for clusters in a similar mass range ($10^{14} \leq M_{\rm 200, mean} / {\rm M}_\odot \leq 10^{15}$), though they do not see the evolution with magnitude cut that we do. Similarly, \citet{Murata2020} find little evolution with magnitude cut, but show that galaxy number counts produce smaller splashback radii than are expected from theoretical calculations.

Other studies employing SZ-selected clusters can give results showing stronger correspondence with the expected dark matter splashback radius when using a similar magnitude cuts \citep{Shin2019,Adhikari2021}.
Both of these studies show measured splashback radii $R_{\rm sp, H} \approx R_{{\rm sp}, i}$ when ensuring that subhaloes are selected with $v_{\rm peak} > 170$ and $150$ km s$^{-1}$ in theoretical simulations respectively.
Here $v_{\rm peak}$ is the highest circular velocity a halo has had over its entire merger history.
Notably, these $v_{\rm peak}$ bounds correspond to a subhalo mass cut of around $M_{\rm H, sub} > 5\times10^{11}$ M$_\odot$ \citep{Reddick2013}. This brings our predictions into alignment, as \citet{Adhikari2021} employs a magnitude cut of $i < -19.87$, making our prediction of $R_{{\rm sp}, M_{\rm H, sub} > 5\times10^{11} {\rm M}_\odot} = 0.92^{+0.21}_{-0.12} R_{\rm 200, mean}$ and $R_{{\rm sp}, i < -20} = 0.87^{+0.05}_{-0.05} R_{\rm 200, mean}$ consistent with theirs for our cluster mass range $10^{14.0} \leq M_{\rm 200, mean} / {\rm M}_\odot \leq 10^{14.5}$.

The magnitude cutoffs also result in a slightly lower, yet consistent, splashback radius than the corresponding stellar mass according to Figure \ref{fig:iband_mcluster}.
This is more pronounced for the lower halo mass ranges.
Taking $i<-18$ to correspond to $M_{*}>10^{9} \rm{M}_{\odot}$, for example, we find that the magnitude cut gives $R_{{\rm sp}, i}/R_{\rm 200,mean}=0.96$, with $R_{{\rm sp}, *}/R_{\rm 200,mean}=0.98$. This minor difference is well within $1\sigma$, and can additionally be explained by slightly different galaxy selections.

\subsection{Redshift evolution}

\begin{figure*}
    \centering
    \includegraphics{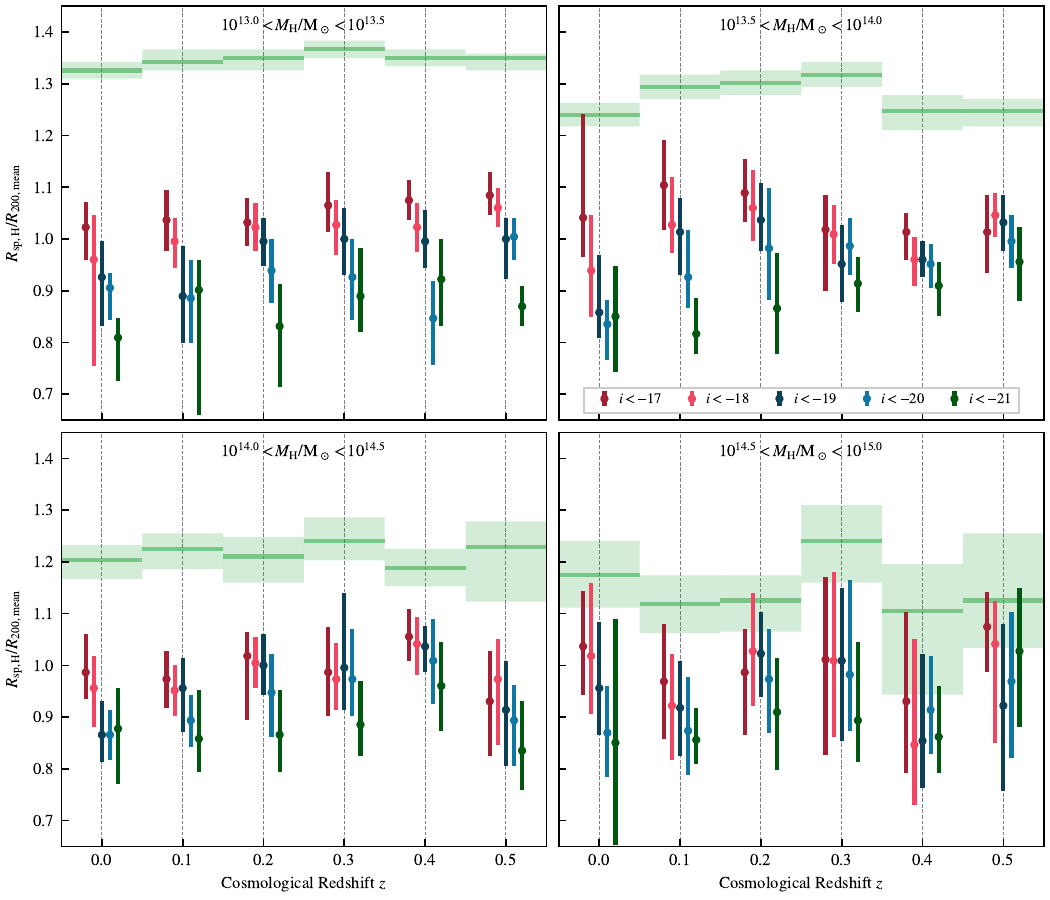}
    \caption{The evolution of splashback radius for the four cluster mass stacks (different panels) as a function of redshift $z$. The green lines indicate the splashback radius as measured from the dark matter profile, with the points showing different absolute magnitude cuts (the same cuts as Fig. \ref{fig:iband_mcluster}). The vertical dashed lines indicate the snapshot redshifts at which the splashback radii are measured, with the points offset around this value as usual for clarity, ordered by the magnitude cut. In general, there is little evolution of the splashback radius (relative to $R_{\rm 200, mean}$), independent of which cut in absolute magnitude is used. In almost all cases, the trend that lower magnitude (brighter) cuts lead to lower measured splashback radii is preserved.}
    \label{fig:splashback_evo}
\end{figure*}

In Figure \ref{fig:splashback_evo}, we consider the evolution of the splashback radius over time.
Each panel shows a different cluster mass, and the points correspond to the same absolute magnitude cuts as in Figure \ref{fig:iband_mcluster}.
As observational surveys frequently stack across a small redshift range, we study the splashback feature at individual snapshots within the simulation for $0.0<z<0.5$ to see the evolution in an extreme case.
The dark matter splashback radius shows very little evolution with redshift for a given halo mass range. The recovered splashback radii measured using a given (rest-frame) $i$-band absolute magnitude cut are again consistent to within their error ranges across our redshift range for all cluster masses.
For most halo mass ranges, we see a continuation of the trend for lower magnitude galaxies to produce smaller splashback radii.
This trend becomes less clear in higher mass haloes at higher redshifts, where our sample size decreases; at $z=0.5$, there are only 13 clusters in our highest mass bin.

These results imply that stacking clusters, even over a large redshift range, should not produce erroneous results as long as a fixed \emph{absolute magnitude} cut-off is used throughout the range, and the appropriate physical $R_{\rm 200, mean}$ is applied to prevent pseudo-evolution.
Observationally, clusters are typically stacked without this scaling.
These studies calculate density profiles using co-moving coordinates to remain consistent across redshifts \citep[e.g.][]{Shin2021}.
Observational samples are usually large enough to allow for significantly narrower cluster mass ranges where clusters will have similar values of $R_{\rm 200, mean}$.
For instance, \citet{Chang2018} found using the Dark Energy Survey that re-scaling using the richness of clusters does not significantly impact their splashback measurements.
Additionally, the exact binning does not significantly alter our results as long as appropriate sampling is achieved (see Appendix \ref{app:binning}).
When considering multiple redshifts, co-moving coordinates will evolve similarly to $R_{\rm200, mean}$, so we expect our qualitative predictions using re-scaled cluster stacks to hold.
Future larger volume simulations will be able to mimic observational methods more closely and produce more accurate predictions.

The underlying galaxy population does not undergo a significant enough evolution between $z=0.5$ and $z=0.0$ to create changes in the splashback radius larger than our measured random errors, at least intrinsically. A potential pitfall here is that these synthetic $i$-band magnitudes are calculated in the rest frame of the galaxies, and as such there would be some significant (by $z=0.5$) movement of light between bands that would need to be considered if attempting to observationally stack galaxies across such a large range. Finally, these results do not include any other observational complications, such as dust extinction of the light, which would further need to be corrected for in observations to create a fair stack. 
\subsection{Star Formation Rates}
\label{sec:results_sfr}

\begin{figure}
    \centering
    \includegraphics{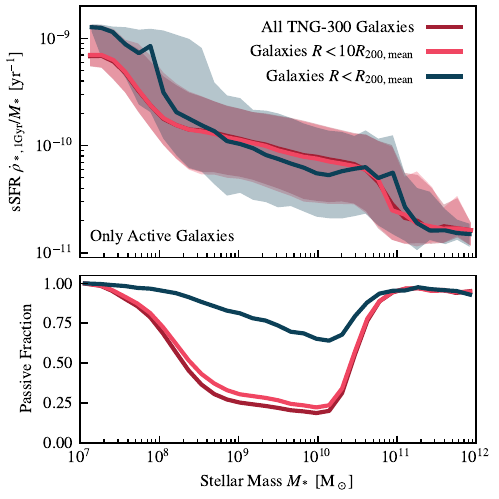}
    \caption{The 1 Gyr-averaged specific star formation rate \citep[sSFR = $\dot{\rho}_* / M_*$, top panel, only showing active galaxies with sSFR $ > 10^{-11}$ yr$^{-1}$, as in e.g.][]{Crain2015} and associated passive fraction of galaxies (bottom panel, using the same definition of active/passive as the upper panel) as a function of their stellar mass. 16-84 percentile ranges are shown as the shaded regions, with the line representing the median in the 32 equally log-spaced bins between $10^7 < M_* / {\rm M}_\odot < 10^{12}$. Three cuts are shown: in red, all galaxies in the TNG-300 simulation; in pink, all galaxies that are included in our sample (i.e. those within $10 R_{\rm 200, mean}$ of a selected cluster; and in navy only galaxies `within' the cluster (with $R < R_{\rm 200, mean}$ of a cluster in our sample).}
    \label{fig:ssfr}
\end{figure}

\begin{figure}
    \centering
    \includegraphics{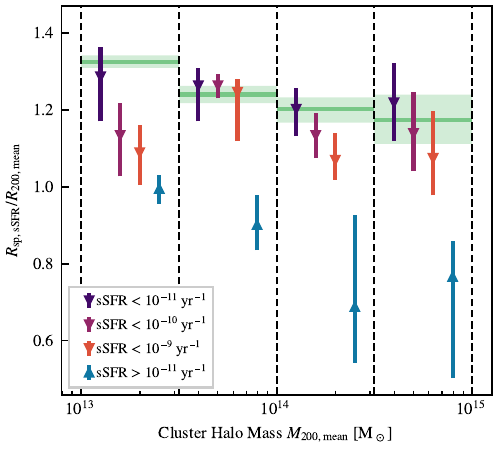}
    \caption{An analogue of Figure \ref{fig:comparemcluster} but now using different cuts in specific star formation rate (sSFR = $\dot{\rho}_* / M_*$) to split the cluster galaxies. Three points (purple to orange, downward arrows) show cuts in sSFR that include all passive galaxies (typically having sSFR $< 10^{-11}$ yr$^{-1}$ classifies a galaxy as being passive) as well as increasing amounts of active galaxies. The blue point (upward arrow) shows the splashback radius measured from only active galaxies. Galaxies with $M_* > 10^8$ M$_\odot$ are included in this analysis, with the star formation rates measured over the past 1 Gyr. Results are qualitatively unchanged when moving to $M_* > 10^9$ M$_\odot$, with the exception of the lowest cluster mass bin where the poor sampling of  galaxies makes fitting impossible.}
    \label{fig:comparemclustersfr}
\end{figure}

Several observational surveys have found that red galaxies produce stronger and more accurate (i.e. closer to the intrinsic dark matter) splashback features than blue galaxies.
Given that the splashback radius aims to separate infalling and collapsed material, this is frequently explained through the quenching of galaxies that have spent time in the cluster environment while blue galaxies are more likely to be on their first infall \citep[e.g.][]{Shin2019,Murata2020,Adhikari2021}.

In IllustrisTNG, red galaxies tend to have a somewhat smaller mass than blue galaxies (e.g. Figure 1 in \citet{Dacunha2021}).
Since lower star formation rates are often associated with a redder colour \citep[e.g.][]{Donnari2019}, this also indicates that quenched galaxies are expected to better trace the theoretical splashback radius. Since we have intrinsic star formation information available, we now attempt to trace the splashback radius with quenched and active galaxies separately.

The top panel of Figure \ref{fig:ssfr} shows the specific star formation rate (sSFR) of passive galaxies in TNG300 averaged over 1 Gyr as a function of stellar mass.
Star formation rates are taken from the catalogues calculated in \citet{Donnari2019}, and are recovered from the birth times of the stars present in the galaxies.
The red line shows the median value for all active galaxies in TNG-300, the pink shows active galaxies in our sample, and the blue shows active galaxies within $R_{\rm 200,mean}$ of a cluster within our sample.
``Active'' is defined as having a specific star formation rate of at least $10^{-11}\ {\rm yr}^{-1}$ \citep{Crain2015}.
The bottom panel of Figure \ref{fig:ssfr} shows the fraction of galaxies in TNG-300 (red), in our sample (pink), and within $R_{\rm200,mean}$ of a sample cluster (blue), with a specific star formation rate less than $10^{-11}\ {\rm yr}^{-1}$.

As shown in the bottom panel, galaxies within the selected clusters are much more likely to be quenched than field galaxies.
For galaxies with a stellar mass of $M_*=10^9\ {\rm M}_{\odot}$, the passive fraction is $\approx0.8$ for galaxies within $R_{\rm200,mean}$ of one of our selected clusters compared to a passive fraction of $\approx0.3$ for all galaxies in the simulation.

Figure \ref{fig:comparemclustersfr} shows the splashback radius computed using samples with various sSFR cutoffs.
Downward facing triangles indicate upper sSFR limits while upward facing triangles indicate lower sSFR limits.
As in Figure \ref{fig:comparemcluster}, the dark matter splashback radius for each halo mass range is shown in green, and the points for various galaxy samples are offset for clarity.
As before, errors are estimated using the $16^{\textrm{th}}$ and $84^{\textrm{th}}$ percentiles of 2048 bootstraps.
Using only passive galaxies (sSFR$<10^{-11}\ {\rm yr}^{-1}$, purple points) in our sample gives good agreement with the dark matter splashback radius.
Increasing the upper limit of sSFR included in our profiles (sSFR$<10^{-10}\ {\rm yr}^{-1}$ and sSFR$<10^{-9}\ {\rm yr}^{-1}$, pink and orange points respectively) increases the number of active galaxies within the sample, which decreases the splashback radius.
Building profiles with only active galaxies (sSFR$>10^{-11}\ {\rm yr}^{-1}$, blue points) gives a very small splashback radius offset nearly $50\%$ from the dark matter splashback radius.
This lower bound includes many galaxies with high sSFR and excludes all quenched galaxies with sSFR$<10^{-11}$, both of which serve to decrease the measured splashback radius.
Notably, we can recover the dark matter splashback radius across our mass range using quenched galaxies, even though in Figure \ref{fig:comparemclusterstellar} we see $R_{\rm sp, *} < R_{\rm sp, DM}$ for $M_* > 10^8$ M$_\odot$.

This is consistent with findings that quenched galaxies are more likely to be collapsed within the cluster.
In the simulations used in \citet{Adhikari2021}, red galaxies had an average halo residence time of more than $t_{\rm res} > 3.2$ Gyrs, while green galaxies had $t_{\rm res} > 2.3$ Gyr.
These populations trace the splashback feature more closely than blue galaxies, which they find have only been in the halo for less than $t_{\rm res} < 1.5$ Gyr and are largely still in-falling.
\citet{Dacunha2021} also found that red galaxies were accreted earlier and better traced the dark matter splashback radius than blue galaxies.

Active galaxies, with a high gas fraction, are affected more by ram pressure stripping than infalling quenched galaxies with lower gas fractions.
This loss of mass and momentum would cause infalling active galaxies to lose more energy and decrease the measured splashback radius.
\citet{Rafieferantsoa2019} found that galaxies that are more massive at infall will stop forming stars more quickly than less massive galaxies, likely due to the loss of gas through dynamical processes like ram pressure stripping.
This stripping does not depend as strongly on host halo mass as dynamical friction does so could help explain how the trends seen in Figures \ref{fig:comparemcluster} and \ref{fig:comparemclusterstellar} persist across cluster mass ranges.
Additionally, the typical quenching time of galaxies found in \citet{Rafieferantsoa2019} is around 2 Gyr (comparable to the crossing time of the cluster), meaning that any active galaxies inside the cluster will have recently fallen in.
With only one or two crossings possible for these active galaxies, they will likely not have had time to virialise with the cluster potential, and hence do not accurately trace out the splashback radius.

Notably, we must use a long-time sSFR to define our galaxy populations because shorter timescales (100 Myr) pick up galaxies that are only recently quenched.
We also note that our results remain qualitatively unchanged when using galaxies with $M_*>10^9\ {\rm M}_{\odot}$ except for the lowest cluster mass bin where there is a poor sampling of galaxies.
This indicates that selecting galaxies that have been in the cluster long enough to virialize is imperative when identifying populations of galaxies that will accurately trace the splashback radius.

This also provides an explanation for why we find a decrease in splashback radius with stellar mass shown in Figure \ref{fig:comparemclusterstellar}.
For galaxies with stellar masses less than $10^{11}$ M$_{\odot}$, there is a decrease in passive fraction with stellar mass.
Thus, although the specific star formation rate of active galaxies decreases with stellar mass (as shown in the top panel of Figure \ref{fig:ssfr}), the \emph{number} of active galaxies, and hence those likely with a short residence time in the cluster, increases with stellar mass.
Increasing stellar mass therefore increases the selection of active galaxies and decreases the splashback radius.

It has been noted in other studies, e.g. \citet{Armitage2018}, that galaxies that have newly entered a cluster are less likely to follow the dark matter potential.
They also find that galaxies tend to lose mass with time spent in the cluster, consistent with \citet{Joshi2017} and \citet{Rhee2017}.
This corresponds with our findings that less massive galaxies, and those that have been quenched for a long time, more closely trace the dark matter splashback radius.
These results together imply that selection criteria for galaxies that result in a more virialized population are likely to result in better tracers for the dark matter splashback radius.

\section{Discussion}
\label{sec:discussion}

From our results, particularly Figures \ref{fig:comparemcluster}, \ref{fig:comparemclusterstellar} and \ref{fig:iband_mcluster}, it is clear that the choice of cut in, or ability to observe, a certain selection of the galaxy population can lead to significant changes in the measured splashback radius from galaxy number counts.
In this section, we consider potential physical reasons that may motivate these results.

\subsection{Evolution in Galaxy Properties}
\label{sec:discussion_galaxy_properties}

\begin{figure}
    \centering
    \includegraphics{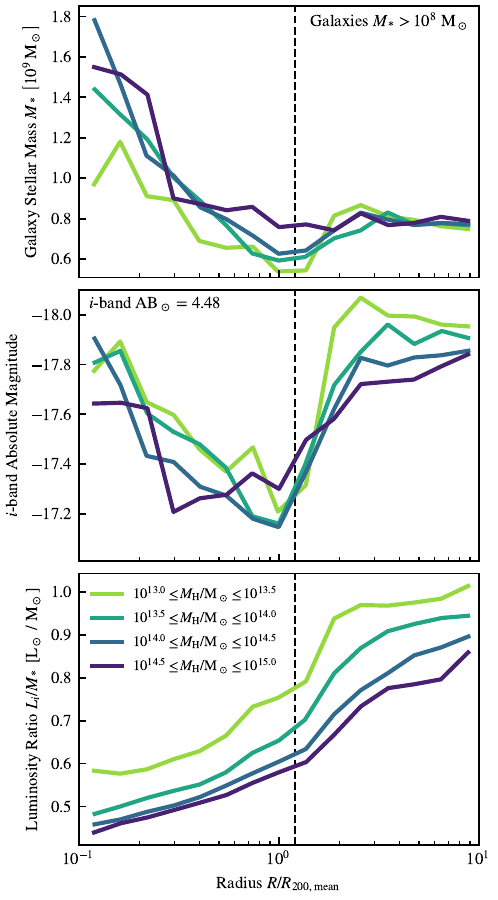}
    \caption{\emph{Top}: Binned median stellar masses of galaxies in 16 equally log spaced bins between $0.1 < R / R_{\rm 200, mean} < 10.0$, for each of the cluster stacks (individual lines). \emph{Center}: Binned median magnitudes of galaxies in the same bins. \emph{Bottom}: Median luminosity-to-mass ratio (calculated per galaxy, and then averaged within the bin) with the absolute $i$-band luminosity of the sun taken to be 4.48. Galaxies are included in the lines if they have $M_* > 10^8$ M$_\odot$. The black dashed line shows a representative dark matter splashback radius for all of our cluster mass ranges, $R = 1.2 R_{\rm 200, mean}$. Galaxies outside the cluster have a uniform brightness, across all cluster masses, and galaxies inside the clusters show an increase in brightness and mass from $R \approx R_{\rm sp}$ to the center. A broadly similar trend is seen in the middle panel when averaging galaxy luminosity instead of $i$-band magnitude.}
    \label{fig:magnitude_comparison}
\end{figure}

Sections \ref{sec:results_galaxies} and \ref{sec:results_luminosity} showed a decrease in $R_{\rm sp}$ with stellar mass and luminosity.
In this section, we examine how these properties evolve for galaxies within and outside clusters.
By developing a better understanding of these properties of galaxies, we may better understand what drives $R_{\rm sp}$ to smaller values.

In Figure \ref{fig:magnitude_comparison}, we show the median stellar masses ($M_*$), magnitudes ($i$-band absolute magnitude) and luminosity ratio (galaxy luminosity in the $i$-band, $L_i$, divided by the galaxy stellar mass, $M_*$) of cluster galaxies as a function of their distance from the cluster centre.
To ensure we have enough stellar particles to model the photometry, we make a stellar mass cut of $M_{*}>10^8\ \rm{M}_{\odot}$ in addition to the magnitude cuts.
The stellar masses of galaxies inside the cluster grow steadily as we move towards the center of the cluster.
Typical galaxies outside the cluster have a mass of around $8\times10^{8}$ M$_\odot$, for all clusters, but this grows to around $1.5\times10^{9}$ M$_\odot$ inside the cluster at $R/R_{\rm 200, mean} \approx 0.1$.
We also see that the median magnitude dips strongly around the splashback radius ($R_{\rm sp} \approx 1.2 R_{\rm 200,mean}$) but recovers to the median outside the cluster at $R/R_{\rm 200, mean} \approx 0.1$.
There is a corresponding decrease in the luminosity ratio, suggesting that these galaxies reach similar luminosities to those outside the cluster because they consist of a larger number of dimmer stars.
This emphasises that the galaxy population within clusters has a distinct stellar population from those outside clusters and that there is a larger abundance of certain types of galaxies within the cluster.
We additionally note that we see a qualitatively similar picture for other bands (e.g. the $g$-band SDSS filter).

The outskirts of clusters will contain a population of backsplash galaxies, which have passed through the cluster in their recent history.
These galaxies lose a significant portion of their dark matter but their stellar population remains similar to when the galaxy initially fell in \citep{Knebe2011}.
This results in a relatively unchanged luminosity distribution compared to galaxies that have not yet entered the cluster, again emphasising that the amount of time spent in the cluster is essential for visibly determining bound haloes.
Since the majority of these backsplash galaxies will fall back in to the halo \citep{Mamon2004, Gill2005, Wetzel2014, Knebe2020}, these backsplash galaxies would ideally be included in the density profiles while the surrounding field galaxies would not.
While we may be able to estimate the fraction of backsplash galaxies for clusters based on observational properties, consistently identifying the backsplash population has proven to be difficult \citep{Oman2013,Haggar2020}.

\begin{figure}
    \centering
    \includegraphics{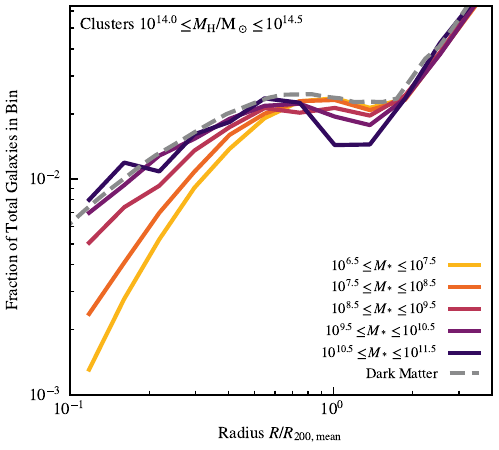}
    \caption{Number density profiles of galaxies in 1 magnitude wide bins for different stellar mass cuts (different lines), shown for the $10^{14.0} < M_H / {\rm M}_{\odot}< 10^{14.5}$ clusters. This figure uses the same binning strategy as Figure \ref{fig:magnitude_comparison}. All lines are normalised to ensure that the outer profile at $R / R_{\rm 200, mean} > 2$ show the same trends. We see that, as the stellar mass decreases, there is a significant under-representation of galaxies in the inner regions. The grey dashed line shows the density profile from the underlying dark matter.}
    \label{fig:mass_profiles}
\end{figure}

In Figure \ref{fig:mass_profiles}, we show the number count profiles (without the included volume of the shell) of galaxies in 1 dex wide bins for main haloes with mass $10^{14.0} \leq M_H/\rm{M}_{\odot} \leq 10^{14.5}$, split by stellar mass.
These profiles show the fraction of the total galaxies in each bin, meaning they are normalised so that the outer number density of galaxies (at $R / R_{\rm 200, mean} > 2$) are uniform
This enables us to clearly see suppression of number counts inside of galaxies. 
As expected, the total number of galaxies decreases for all magnitude bands towards the centre, where the radial bins are naturally smaller in volume.

\citet{Adami1998} also noted that there are more luminous galaxies closer to the cluster centre in observations of 40 clusters.
They also found that these galaxies have a lower velocity dispersion than less luminous galaxies and that this relationship holds only for elliptical type galaxies.
As discussed in Section \ref{sec:results_sfr}, this indicates that a more virialized population of galaxies best represents the gravitational potential of a cluster.
Figure \ref{fig:iband_mcluster} indicates that brighter galaxies trace $R_{\rm sp}$ less accurately, but the galaxy population included in these results is made up of both ellipticals and spirals.
Thus, it is not sufficient to select galaxies based solely on magnitude to accurately measure the splashback radius.
Since luminosity appears to vary significantly with radius, it may also be possible to more accurately trace the dark matter potential with a luminosity-weighted density profile as in e.g. \citet{Bianconi2021}.

Starting at $R/R_{\rm200,mean}\approx2$, we see a dip in the number of galaxies with high masses, but this quickly recovers to trace the dark matter density profile. The lower mass galaxies generally drop in abundance, relative to the expected value based upon the dark matter, the further into the centre of the cluster we go. The abundance of galaxies is lower with lower masses, corroborating the increase in stellar mass that we saw in Figure \ref{fig:magnitude_comparison}. It appears that this increase in mean stellar mass is not due to the growth of galaxies, but the destruction (and merger with the central object) of lower mass galaxies.

The lack of overabundance of massive galaxies relative to the dark matter, and the low luminosity ratios in the centre of the cluster, imply that the galaxies in the cluster are not undergoing significant star formation due to their new, denser environment.
This confirms that the cluster galaxies population differ in intrinsic properties influenced by their environment, so selecting galaxies based on these properties will likely influence the measured density profiles and, therefore, measurements of the splashback radius.

\subsection{Galaxy Abundances}
\label{sec:discussion_galaxy_abundances}

That the density and concentration of galaxies in clusters differs from what would be expected based upon number counts in the field is previously known \citep[e.g.][]{Kaiser1984,Scoccimarro2000,Budzynski2012}.
This is expected if galaxies are systemically changed as they enter a cluster and therefore differ from the field galaxy population.
If galaxies are stripped of their mass as they fall in, for example, there will be fewer high mass objects relative to the background density.
\citet{Armitage2018} find that galaxies that have been in clusters for longer tend to have a higher stellar-total mass ratios, which is consistent with diffuse dark matter being easily from galaxies as they orbit a cluster. 
Additionally, lower mass galaxies have a lower survival time within clusters than their higher-mass counterparts, and as such we would expect some evolution in the abundance of galaxies with mass \citep{Chua2017, Bahe2019}.

A change in galaxy abundance would not have an impact on the true splashback radius of the cluster, which is determined by the potential, but may impact our measurement of it using number density profiles. In this section we investigate how accurately the galaxies trace the underlying density profile, by comparing the ratio of their density inside of the cluster to the outside relative to the dark matter. We then investigate how this inaccuracy would impact our splashback measurement.

\begin{figure}
    \centering
    \includegraphics{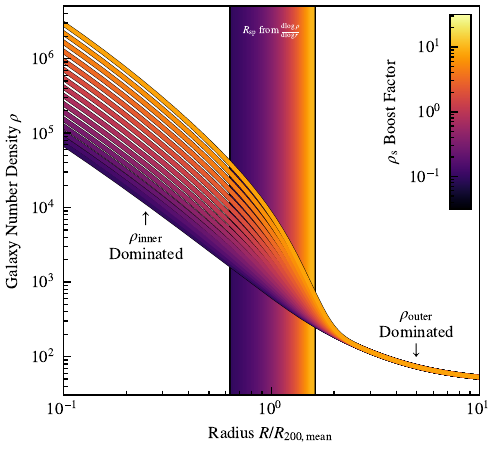}
    \caption{Shows the impact of varying $\rho_{\rm s}$ (in Eqn. \ref{eq:density}) by a `boost factor' on the density profile (foreground; 16 examples shown), and the resultant change in the measured splashback radius (background gradient), based on the intrinsic dark matter profile for the $10^{14.0} < M_H / {\rm M}_\odot < 10^{14.5}$ clusters. Decreasing the central density moves the measured splashback radius in, and vice versa. The cross-over point between these two profiles is what sets the splashback radius, meaning that any changes in the abundances of galaxies inside clusters (relative to the intrinsic halo population) due to morphological changes can impact the splashback radius measurements.}
    \label{fig:densityratiovisual}
\end{figure}

\begin{figure}
    \centering
    \includegraphics{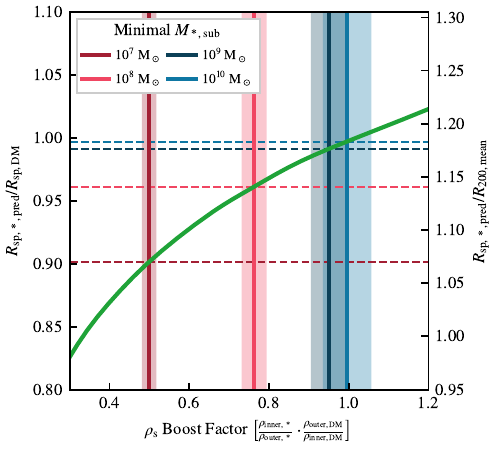}
    \caption{Shows the background gradient from Figure \ref{fig:densityratiovisual} as the green line, showing how the predicted splashback radius (horizontal dashed lines) should change if it was determined only by boosts or suppressions in the cluster density. Each vertical line shows the measured boost factor (see text) for the four stellar mass cuts in Figure \ref{fig:comparemclusterstellar}. The shaded regions show 16-84 percentile ranges based upon the selection A bootstraps. Here, the trend is in entirely the wrong direction, with the simple density argument leading to larger measured splashback radii for cuts that only include more massive galaxies.}
    \label{fig:densityrationumeric}
\end{figure}

In our analysis, this lower abundance of galaxies in clusters would manifest as $\left(\rho_{\rm inner}/\rho_{\rm outer}\right)_{\rm DM}>\left(\rho_{\rm inner}/\rho_{\rm outer}\right)_{\rm gal}$ with $\rho_{\rm inner}$ and $\rho_{\rm outer}$ as defined in Equation \ref{eq:density}. The two major components of this model, $\rho_{\rm inner}$ and $\rho_{\rm outer}$, fit to material currently inside the halo and an infalling component respectively. $f_{\rm trans}$ acts to sharpen the transition between the two components, as is commonly required in practice, and notably the truncation radius $r_{\rm t}$ need not coincide with the measured position of the splashback radius $R_{\rm sp}$. To study the impact of this systematic on our measurement of the splashback radius, we define a ``boost factor"
\begin{equation}
    \indent \rho_{\rm s,boost} = \frac{\rho_{\rm inner,x}}{\rho_{\rm outer,x}} \cdot \frac{\rho_{\rm outer,DM}}{\rho_{\rm inner,DM}}
\end{equation}
as the relative overdensity of a halo component (e.g. ${\rm x} = {\rm gal}$) in the inner part of the cluster compared to the overdensity of the dark matter. Taking 
\begin{align}
    \indent&\rho_{\rm inner} = \rho\left(0.2R_{\rm200,mean}<R<0.5_{\rm200,mean}\right) \nonumber\\
    &\rho_{\rm outer} = \rho\left(2R_{\rm200,mean}<R<5R_{\rm200,mean}\right)\ ,
\end{align}
we explore the impact of varying $\rho_{\rm s}$ from Equation \ref{eq:density} by a boost factor. We do not vary the parameters of $f_{\rm trans}$ as this factor of order unity does not cause significant qualitative changes in the density profile and would unnecessarily complicate the discussion that follows. Figure \ref{fig:densityratiovisual} shows the density profile for 16 example boost factors varying from 0.1 to 10 for the dark matter profile of clusters with $10^{14.0}<M_H/\rm{M}_{\odot}<10^{14.5}$.
In particular, higher boost factors, i.e. increasing the inner density of the profile, leads to a larger splashback radius while still allowing the profiles to converge to the background density at large radii.

We see a strong trend for more massive galaxies to have a smaller splashback feature, so we test the ability of the boost factor to explain this.
For a given stellar mass cut, we calculate the boost factor of that profile.
We then predict the expected splashback radius for that boost factor using the background gradient from Figure \ref{fig:densityratiovisual} and show our results in Figure \ref{fig:densityrationumeric}.
We find that this predicts an increasing splashback radius for higher stellar mass cuts since lower mass galaxies are underrepresented relative to higher mass galaxies within the cluster.
However, this is the opposite of the trend shown in Figure \ref{fig:comparemclusterstellar}.
Thus, differences in abundances cannot explain the trends we see for decreasing $R_{\rm sp}$ for higher galaxy masses, though they may still influence our results.

With no difference in the abundance of higher mass galaxies, this leaves us with the conclusion that the major factor leading to smaller measured splashback radii for this population is that they are more centrally concentrated at radii around the splashback radius. This conclusion is supported by Fig. \ref{fig:magnitude_comparison}, showing that the mean mass of galaxies increases closer to the cluster centre, and Fig. \ref{fig:mass_profiles} that shows a gradient change at lower radii monotonically with galaxy stellar mass.

\section{Conclusions}
\label{sec:conclusions}

Constraining the physical extent of the influence of galaxy clusters is key to understanding differences between cluster and field galaxies. \citet{More2015} suggested that the `splashback radius', $R_{\rm sp}$ provides a significantly improved alternative for denoting the cluster boundary than overdensity based criteria (e.g. $R_{\rm 200, mean}$). However, measuring $R_{\rm sp}$ is not easy in observations, as it requires finding a caustic in the underlying total matter density profile of the cluster. \citet{O'Neil2021} showed that  $R_{\rm sp}$ can be accurately determined from the number counts of cluster galaxies when many clusters in a similar mass range are stacked.

In this paper, we investigated how various cuts in the cluster galaxy population can impact measurements of $R_{\rm sp}$ based upon galaxy number counts by using the IllustrisTNG-300-1 cosmological galaxy formation simulation. We calculated $R_{\rm sp}$ for isolated galaxy clusters in the mass range $10^{13} \leq M_{\rm 200, crit} / {\rm M}_\odot \leq 10^{15}$, stacked in 0.5 dex wide bins. To extract $R_{\rm sp}$, we fit the physically motivated density profile model from \citet{Diemer2014} directly to the logarithmic derivative of the density profile $\rho$ as a function of radius $r$, $\mathrm{d}\log\rho / \mathrm{d}\log r$, following \citet{O'Neil2021} and Appendix \ref{app:fitting}. $R_{\rm sp}$ was identified at the radius at which $\mathrm{d}\log\rho / \mathrm{d}\log r$ was at a minimum.

The main results from this paper can be summarised as follows:
\begin{itemize}
    \item Based upon minimum cuts in subhalo mass $M_{\rm H, sub}$ from $10^9$ M$_\odot$ to $10^{11}$ M$_\odot$ and minimum cuts in galaxy stellar mass $M_*$ from $10^7$ M$_\odot$ to $10^{10}$ M$_\odot$, we determined that only the smallest galaxies ($M_{\rm H, min, sub} = 10^9$ M$_\odot$) in the simulation could trace the intrinsic dark matter $R_{\rm sp}$ when mass cuts alone were used. More massive galaxies trace significantly smaller values of $R_{\rm sp}$ with, for example, $M_{\rm H, sub} > 10^{11}$ $M_\odot$ tracing a splashback radius that was around 25\% smaller across our mass range (Figures \ref{fig:comparemcluster} and \ref{fig:comparemclusterstellar}).
    \item By employing synthetic galaxy luminosities within relevant SDSS Camera Response Function bands, we showed that the expected trends for brighter (higher mass) galaxies tracing smaller values of $R_{\rm sp}$ were repeated, though we were unable to find any reasonable absolute magnitude cut that would enable the galaxy number counts to align with the predictions from the dark matter (Figure \ref{fig:iband_mcluster}). We saw an offset of at least 20\% from the measured dark matter $R_{\rm sp}$ even when using an untenable cut in $i$-band absolute magnitude of $i < -17$.
    \item We were able to reconcile observations from both optically-selected \citep{Baxter2017, Zurcher2019, Murata2020, Gonzalez2021}, that typically find $R_{\rm sp}$ measured from number counts to be smaller than theoretical predictions (directly measured from N-body simulations), and SZ-selected clusters \citep{Shin2019, Adhikari2021}, that typically find alignment between the two measurements, by demonstrating differences in their cluster galaxy selection functions. \citet{Shin2019} and \citet{Adhikari2021} compare their observations against splashback radii calculated from only the most massive substructures in simulations (and hence, from our Figure \ref{fig:comparemcluster}, a smaller `theoretical' splashback radius), bringing their observationally derived $R_{\rm sp}$ into alignment with simulations. We demonstrated that our results from TNG-300 produce measurements of $R_{\rm sp}$ from cuts in $i$-band magnitude that are consistent with observations from a wide range of studies.
    \item In Figure \ref{fig:splashback_evo} we showed that there is little evolution in $R_{\rm sp}$ measured from the intrinsic dark matter, or from galaxy number counts (when using a fixed $i$-band cut in \emph{absolute magnitude}) across a redshift range of $0.0 \leq z \leq 0.5$. This implies that studies are able to stack clusters across a wider redshift range than is currently used \citep[e.g. $0.1 \leq z \leq 0.33$][]{Baxter2017}.
    \item In Figure \ref{fig:comparemclustersfr}, following suggestions from \citet{Murata2020}, \citet{Adhikari2021}, \citet{Dacunha2021}, and others that red galaxies trace a more accurate $R_{\rm sp}$ than blue galaxies, we consider how cuts in 1 Gyr averaged specific star formation rate (sSFR) impact our measurement of $R_{\rm sp}$ (we note that 100 Myr averaged sSFR were not able to produce a reliable result). We found that quenched galaxies (sSFR $< 10^{-11}$ yr$^{-1}$) were able to reproduce the dark matter measurement consistently across our mass range to within $1\sigma$. Active galaxies, however, trace $R_{\rm sp}$ up to 30\% smaller than expected. We showed that steadily including more active galaxies (by increasing our threshold in sSFR) ensured that our measurement of $R_{\rm sp}$ decreased monotonically, further increasing confidence in our result. 
    \item We showed that cluster galaxies are typically more massive than those in the field, but less bright, suggesting a significant divergence between the properties of field and cluster galaxies (Figure \ref{fig:magnitude_comparison}), and that this change in galaxy properties occurs around the measured dark matter splashback radius. We showed in Figure \ref{fig:mass_profiles} that this change in average galaxy mass was due to an underabundance of the smallest galaxies with $M_* < 10^{9.5}$ M$_\odot$. Figure \ref{fig:densityratiovisual} suggests that this simple change in abundance would lead to significant evolution in the measured $R_{\rm sp}$, providing a possible explanation for our results in Figure \ref{fig:comparemclusterstellar}. Our Figure \ref{fig:densityrationumeric} showed that the changes in abundance of cluster galaxies relative to their field cousins of a similar mass could only account for an offset of around 10\% in $R_{\rm sp}$, and that this offset was negligible for the highest mass galaxies. These results suggest that a simple cut in mass, or magnitude, cannot bring the dark matter measurement and number count measurement of $R_{\rm sp}$ into alignment.
\end{itemize}

Our findings suggest that differences between observation and theory suggested in numerous papers may be explained by subtle differences in galaxy selection functions. As we find such a strong evolution of the splashback radius with mass and brightness cut, any differences in the mass of substructure used to reconstruct the splashback radius in simulations, or differences in chosen absolute magnitude cut when stacking, may lead to significant apparent divergences between observation and theory.

A number of recent studies \citep[e.g.][]{Adhikari2021} have suggested that the major determinant in how well a given galaxy population reconstructs the underlying gravitational potential is their residence times within the cluster. Our cut in sSFR, selecting quenched galaxies, goes some way to constructing a viable sample, but more research is needed to viably select a population of galaxies that can accurately reproduce the dark matter splashback radius by preferentially selecting those with long residence times in a way that is observationally viable.

\section{Acknowledgements}

The authors thank the anonymous referee and Joop Schaye for comments that improved the quality of the manuscript.
The authors would like to thank David Barnes for his contribution to this work before a career move.  We also thank Susmita Adhikari, Michael McDonald and Dhayaa Anbajagane for helpful feedback.
Some of the computations in this paper were run on the FASRC Cannon cluster supported by the FAS Division of Science Research Computing Group at Harvard University. Some of the computations were performed on the Engaging cluster supported by the Massachusetts Institute of Technology. MV acknowledges support through NASA ATP 19-ATP19-0019, 19-ATP19-0020, 19-ATP19-0167, and NSF grants AST-1814053, AST-1814259, AST-1909831, AST-2007355 and AST-2107724.

\section{Data Availability}

The data underlying this article are available in Zenodo at \url{https://dx.doi.org/10.5281/zenodo.5898396}.
All data in this article were reduced from the publicly available TNG300-1 simulation data available at \url{https://tng-project.org} \citep{Nelson2019}.

\bibliographystyle{mn2e}
\bibliography{bibliography}

\appendix

\section{Binning In a Data Poor Environment}
\label{app:binning}

\begin{figure}
    \centering
    \includegraphics{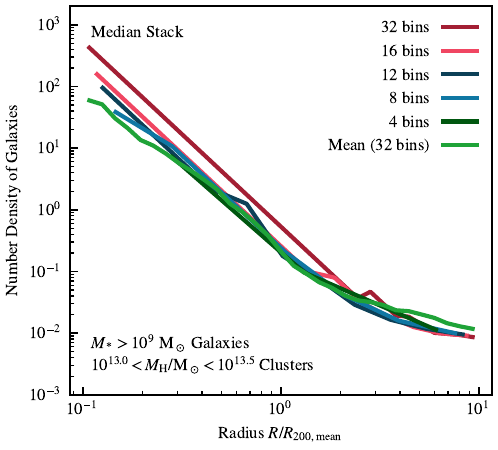}
    \caption{Shows the impact of stacking binned data using the median, for various bin numbers. Here we use all cluster galaxies with $M_* > 10^9$ M$_\odot$ and all clusters with mass $10^{13} < M_{\rm H} / {\rm M}_\odot < 10^{13.5}$. To construct the median, clusters with zero galaxies in each bin must be rejected. Bins are chosen to be equally logarithmically spaced between $0.1 < R / R_{\rm 200, mean} < 10.0$. The light green line shows the red line from Figure \ref{fig:mean_stack}, representing a mean stack of clusters using 32 logarithmically spaced bins.}
    \label{fig:median_stack}
\end{figure}

In \citet{O'Neil2021}, median stacking was used when combining all clusters in a given mass range. Figure \ref{fig:median_stack} shows how the cluster density profile, for galaxy number density, changes when using different numbers of bins if using median stacking. Here, profiles with zero galaxies in a bin must be rejected, and the median calculated from the remaining sample, or no profile can be calculated (the median is usually zero in these bins). The abundance of clusters with a single galaxy in each bin, particularly for the 32 and 16 bin case, leads to the vast majority of bins following a density line representing a single galaxy in the bin. The transition to the median representing multiple galaxies in each bin (which occurs around $R/R_{\rm 200, mean}\approx 2$, coinciding with the typical radius of the splashback feature) can then lead to a false positive detection for the splashback feature due to the rapid gradient change.

It is not plausible to simply decrease the number of bins in the central region due to the large number of parameters in Equation \ref{eq:density_derivative}, and the need to construct a numerical density gradient. For instance, $\rho_{\rm inner}$ has three parameters and is typically only constrained by radii smaller than $R \approx 0.6 R_{\rm 200, mean}$. As Figure \ref{fig:median_stack} shows, 8 or fewer bins for the whole radii range is required to provide an adequate sampling of galaxies in the inner bins. Here, we then run into discretisation errors; as the number of galaxies in the bin must be a (small) integer, there can be a significant error simply because the `true' density is between integers (e.g. 1.5, but the median is given as 1 or 2, a $50\%$ error).

\begin{figure}
    \centering
    \includegraphics{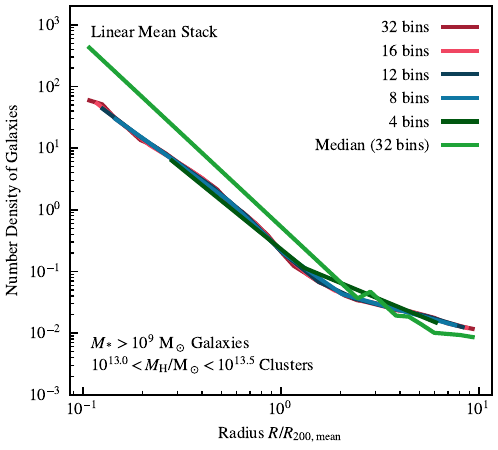}
    \caption{An analogue to Figure \ref{fig:median_stack}, but now using a linear mean to stack clusters in a given bin. Here, again, we use all cluster galaxies with $M_* > 10^9$ M$_\odot$ and all clusters with mass $10^{13} < M_{\rm H} / {\rm M}_\odot < 10^{13.5}$. This leads to a consistent result for the density independent of the number of bins. The light green line shows the red line from Figure \ref{fig:median_stack}, representing a median stack of clusters using 32 logarithmically spaced bins between $0.1 < R / R_{\rm 200, mean} < 10.0$.}
    \label{fig:mean_stack}
\end{figure}

Figure \ref{fig:mean_stack} shows a similar figure but now using different numbers of bins to stack the clusters in the sample with a linear mean of galaxies in each bin. This is a true `stack', where now each bin contains the information from all possible cluster galaxies, with the line being normalised by the number of clusters in the sample. The results are now almost entirely independent of the number of bins used, and as such this is the approach that we take for stacking the clusters in this work. The benefits of median stacking, where outliers are prevented from impacting the sample average, are lost when it is impossible to accurately reconstruct the true stacked density. Our definition ensuring that the clusters are isolated (see \S \ref{sec:methods_gal_def}) prevents extreme sample variance well enough for this to not be a significant problem.

\section{Fitting Profiles}
\label{app:fitting}

In this appendix we explore a number of potential fitting methods for the galaxy profiles (both parametric and non-parametric), and fully describe our bootstrapping strategy.

\subsection{Bootstrapping Errors on the Splashback Radius}

All splashback radius measurements are constructed from 2048 bootstraps amongst the individual clusters making up the stacked sample. Using these bootstraps, we take the median to represent the predicted value, and the 16-84 percentile range as the $1\sigma$ error on our measurement.

We use two levels of bootstrapping, as we require errors on the stacked gradient profiles to improve the quality of fits. To do this, the same number of clusters in each bin are randomly resampled with replacement for each top-level bootstrap (selection A). We then perform another level of bootstrapping, again resampling selection A with replacement to construct 32 selection B samples. The selection B samples have their profiles stacked using a linear mean sum. The gradient is calculated using a second order central differencing scheme \citep[using {\tt np.gradient}][]{Harris2020} with first order edges. The 32 selection B gradients and profiles are then used to construct the mean and standard deviation in each bin to represent the prediction and error respectively, and the methods (described below) are used to construct a fit and hence splashback radius for each unique selection A sample. This procedure is required as it is imperative to have representative errors on each point for the non-parametric fitting schemes described below.

The underlying reason why this two level approach to bootstrapping is required is because of our use of the numerical gradient in the fitting process. It would, in principle, be possible to construct errors on individual density bins (e.g. the error on an individual density point could reasonably be represented as the standard deviation amongst the stack), but propagating this through the numerical gradient scheme is not possible. This propagation either leads to overestimated errors (when the error is taken as the gradient from the top of one errorbar to the bottom of the adjacent errorbar), or unrepresentative errors (when using the upper bound on the density and lower bound on the density to `bracket' the mean).

\subsection{Parametric Fits}

For the parametric fits, we use the physically motivated cluster density profile from \citet{Diemer2014} and its analytic derivative, as used in \citet{O'Neil2021}. For the case of fitting to the galaxy profiles, we treat the mean density of the universe, $\rho_{\rm m}$, as a free parameter (as this will change in unpredictable ways with our various cuts in galaxy mass and luminosity).

In all cases, we first fit the three components of Equation \ref{eq:density} separately. We first fit the density profile $\rho_{\rm inner}$ to the inner data at radii $0.2 < r / R_{\rm 200, mean} < 1.0$, and simultaneously fit the outer profile $\rho_{\rm outer}$ at radii $r / R_{\rm 200, mean} > 2.0$. We provide limits on the values of the free parameters to prevent overflows in the fitting process, with $10^{-10} \rho(r = 0.1 R_{\rm 200, mean}) < \rho_{\rm s}$, $0.01 R_{\rm 200, mean} < r_{\rm s} < 10.0 R_{\rm 200, mean}$, $0.0 < \alpha < 10.0$, $0.1 \rho(r = 10 R_{\rm 200, mean}) < \rho_{\rm m} < 10.0 \rho(r = 10 R_{\rm 200, mean})$, $1.0 < b_e < 10.0$, and $0.0 < S_e < 10.0$.

The free parameters in the density functions are then frozen as we fit the transitional profile $f_{\rm trans}$ using all data points. Here, we ensure that the free parameters $0.01 R_{\rm 200, mean} < r_{\rm t} < 10.0 R_{\rm 200, mean}$, $0.0 < \beta < 10.0$, and $0.0 < \gamma < 5.0$. Functions are fit using the {\tt curve\_fit} function provided in the {\tt scipy} {\tt python} library \citep{Scipy2020}.

The individually fit parameters are then used as an initial parameter value when fitting all simultaneously to all data, whether this is to the density or gradient profile.

\subsection{Non-Parametric Fits}

To investigate whether it is possible to reconstruct the splashback radius (through a fit) from the data without relying on any underlying physical knowledge, we turn to non-parametric fitting through Gaussian Process Regression (GPR). For our GPR fits, we use the {\tt python} library {\tt george} \citep{Ambikasaran2015}.

To fit the profiles and gradients (separately), we use a one-dimensional exponential squared kernel ({\tt ExpSquaredKernel}), and find the maximum liklehood parameters for this kernel by minimising the marginalised log-liklehood using the  Broyden–Fletcher–Goldfarb–Shanno (BFGS) algorithm, implemented in the {\tt scipy} library \citep{Scipy2020}.

\subsection{Example Profile}

\begin{figure}
    \centering
    \includegraphics{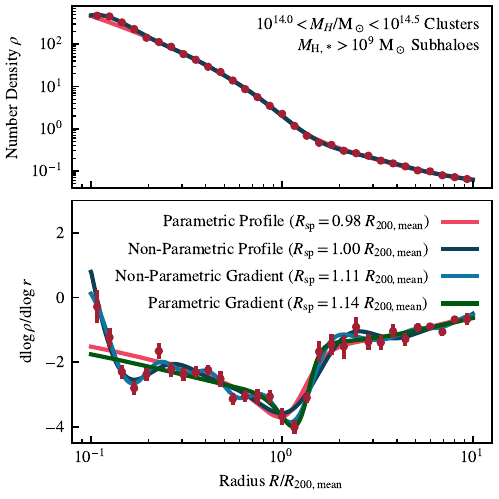}
    \caption{Fitting an example profile with four methods. The top panel shows the number density profile of galaxies, and the bottom shows the differential profile. The red points show the profile constructed for a selection A bootstrap, with the mean and errors constructed from the selection B bootstrap (see text for details) representing one standard deviation of scatter. When fitting the density profile directly we over-smooth the splashback feature and underestimate the splashback radius relative to those that fit to the gradient directly.}
    \label{fig:parametricvsnonparametric}
\end{figure}

Figure \ref{fig:parametricvsnonparametric} shows a (random) example profile, one selection A bootstrap, from the $10^{14} < M_H / {\rm M}_\odot < 10^{14.5}$ cluster sample, with a cut in galaxy stellar mass meaning only galaxies with a stellar mass $M_* > 10^9$ M$_\odot$ are included in the profile. The lower panel shows the differential profile, with four different fitting methods applied, and clearly shows how methods that fit directly to the differential profile are better able to fit the sharp dip in ${\rm d} \log \rho / {\rm d} \log r$ and accurately recover the splashback radius. Note that when fitting the density profile, we always fit directly to the logarithm of the density $\log\rho$, following \citet{O'Neil2021}.

For all fitting methods, we compute the fit profile (or gradient) at 1024 equally log-spaced intervals between $0.1 R_{\rm 200, mean}$ and $10.0 R_{\rm 200, mean}$. If the fit is to the profile, this is then numerically differentiated using the same algorithm as is applied to the data, and the splashback radius for this bootstrap is then taken to be the radius at which the fit differential profile ${\rm d}\log\rho / {\rm d} \log r$ is at a minimum.

\subsection{Comparing Splashback Radii}

\begin{figure*}
    \centering
    \includegraphics{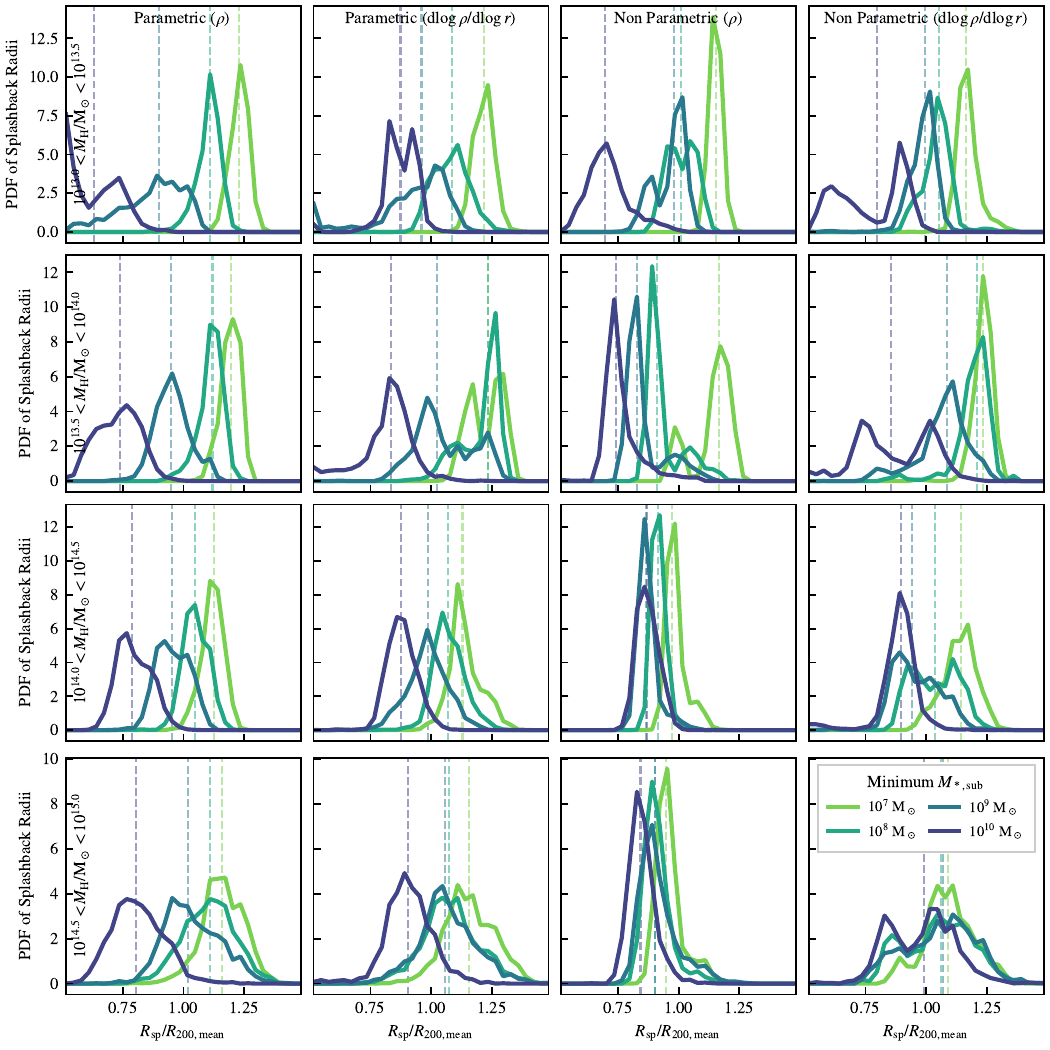}
    \caption{Splashback radii distributions for four different fitting methods (columns): the parametric Equation \ref{eq:density}, the parametric differential Equation \ref{eq:density_derivative}, the non-parametric fitting method described in the text using Gaussian Process Regression, and a non-parametric fitting method using the gradient directly rather than the original profile. Each row shows the different cluster mass cuts, and each line colour shows the different cuts in galaxy stellar mass. We see that the broad trends of the splashback radius decreasing as we increase the cut in galaxy stellar mass is reproduced irrespective of fitting method.}
    \label{fig:measuredsplashbackradii}
\end{figure*}

In Figure \ref{fig:measuredsplashbackradii}, we compare distributions of splashback radii from all four fitting methods when changing the minimum galaxy stellar mass, across our entire cluster mass range. Notably, we see clear bell-shaped profiles across the whole mass range for all fitting methods, with this being the clearest for the parametric gradient fit (used in the rest of the paper).

The non-parametric gradient fitting method tends to produce bimodal distributions in splashback radii. On close inspection of the fit profiles, this is due to it over-fitting noise in the poorly sampled inner regions of the halo, which does not occur with the parametric methods. This shows that our prior of having a physical understanding of the density profile structure is of significant advantage.

All four methods clearly show the progression of the splashback radius decreasing as the minimum stellar mass for galaxies included in the profile is increased, showing that our main results are robust to the fitting method employed.

\label{lastpage}

\end{document}